\documentclass[12pt,epsf,qsymbols]{article}
\pdfoutput=1
\usepackage[utf8]{inputenc}
\usepackage[normalem]{ulem}
\usepackage[english]{babel}%,frenchb
\usepackage{tabularx}
\usepackage{array}
\usepackage{graphics}
\usepackage[pdftex]{graphicx}
\usepackage{psfrag}
\usepackage{epsfig}
\usepackage{amsmath}
\usepackage{amssymb}
\usepackage{bbold}
\usepackage{setspace}
\usepackage{rotating}
\usepackage{colortbl}
\usepackage{longtable}
\usepackage{slashed}
\usepackage{braket}
\usepackage{lineno}
\makeatletter
\usepackage{textcomp}
\usepackage[usenames,dvipsnames]{xcolor}
\usepackage{relsize}
\usepackage{cite}

\usepackage{bbm}
\usepackage{bm}
\usepackage{enumerate}
\usepackage{amsmath}
\usepackage{verbatim}

\usepackage{hyperref}
\hypersetup{colorlinks,citecolor=nicegreen,linkcolor=niceblue}
\hypersetup{colorlinks=true}

\usepackage{amssymb}% http://ctan.org/pkg/amssymb
\usepackage{pifont}% http://ctan.org/pkg/pifont

\setlongtables

\newcommand{\Rkst}{{R_{K^\ast}}}
\newcommand{\cb}{{\mathcal B}}
\newcommand{\bea}{\begin{eqnarray}}
\newcommand{\eea}{\end{eqnarray}}
\newcommand{\beq}{\begin{equation}}
\newcommand{\eeq}{\end{equation}}
\newcommand{\ec}{\end{center}}
\newcommand{\bc}{\begin{center}}

\newcommand{\tev}{{\rm TeV}}
\newcommand{\gev}{{\rm GeV}}

\newcommand{\pdir}{p\kern -5.2pt\raise 0.2ex\hbox {/}}

\newcommand{\vdir}{v\kern -5.75pt\raise 0.15ex\hbox {/}}
\newcommand{\kdir}{k\kern -5.75pt\raise 0.15ex\hbox {/}}
\newcommand{\epsdir}{\epsilon\kern -5.0pt\raise 0.15ex\hbox {/}}
\newcommand{\bvdir}{\bar{v}\kern -5.75pt\raise 0.15ex\hbox {/}}
\newcommand{\Ddir}{D\kern -7.75pt\raise 0.20ex\hbox {/}}
\newcommand{\Adir}{A\kern -7.75pt\raise 0.20ex\hbox {/}}
\newcommand{\ldir}{l\kern -5.0pt\raise 0.2ex\hbox{/}}
\newcommand{\varepsdir}{\varepsilon\kern -5.5pt\raise 0.15ex\hbox{/}}

\newcommand{\nn}{\nonumber}
\newcommand{\mrm}[1]{\mathrm{#1}}

\newcommand{\re}[0]{\mrm{Re}}
\newcommand{\im}[0]{\mrm{Im}}
\makeatother

\definecolor{niceblue}{rgb}{0.15,0.15,0.6}
\definecolor{nicegreen}{rgb}{0.1,0.5,0.1}
\definecolor{Red}{rgb}{1.,0.,0.}

\definecolor{Green}{rgb}{0.2,.7,0.2}

\begin{document}
\unitlength = 1mm

\thispagestyle{empty} 
\begin{flushright}
\begin{tabular}{l}
{\tt \footnotesize LPT 17-20}\\
\end{tabular}
\end{flushright}
\begin{center}
\vskip 3.4cm\par
{\par\centering \textbf{\LARGE  
\Large \bf A leptoquark model to accommodate  \\ 
\vskip .3cm 
\Large \bf $R_K^\mathrm{exp} <R_K^\mathrm{SM} $ and $R_{K^\ast}^\mathrm{exp}< R_{K^\ast}^\mathrm{SM}$}}\\
\vskip 1.2cm\par
{\scalebox{.85}{\par\centering \large  
\sc Damir~Be\v{c}irevi\'c$^a$ and Olcyr~Sumensari$^{a,b}$}
{\par\centering \vskip 0.7 cm\par}
{\sl 
$^a$~Laboratoire de Physique Th\'eorique (B\^at.~210)\\
CNRS and Univ. Paris-Sud, Universit\'e Paris-Saclay, 91405 Orsay cedex, France.}\\
{\par\centering \vskip 0.25 cm\par}
{\sl 
$^b$~Instituto de F\'isica, Universidade de S\~ao Paulo, \\
 C.P. 66.318, 05315-970 S\~ao Paulo, Brazil.}\\

{\vskip 1.65cm\par}}
\end{center}

\vskip 0.85cm
\begin{abstract}
We show that a modification of the model with a low energy scalar leptoquark state carrying hypercharge $Y=7/6$ allows to accommodate 
both $R_K<1$ and $R_{K^\ast}<1$, through loop effects, consistent with recent observations made at LHCb. We describe details of the model, compute the relevant Wilson coefficient and, after 
discussing a number of constraints, we examine the phenomenological consequences of the model. The bounds on the lepton flavor violating decay rates, induced by this model, 
include $\cb(Z\to\mu\tau) \lesssim \mathcal{O}(10^{-7})$, and $\mathcal{B}(B\to K\mu\tau) \lesssim \mathcal{O}(10^{-9})$. We also comment on the interpretation of the bounds on the leptoquark mass
obtained from the direct searches at LHC. 
\end{abstract}
\newpage
\setcounter{page}{1}
\setcounter{footnote}{0}
\setcounter{equation}{0}
%%%%%%%%%%%%%%%%%%%%%%%%%%%%%%%%%%%%%%%%
\noindent

\renewcommand{\thefootnote}{\arabic{footnote}}
%\linenumbers

\setcounter{footnote}{0}

%\tableofcontents

\newpage

%%%%%%%%%%%
%%%%%%%%%%%
%%%%%%%%%%%
\section{Introduction}
\label{sec:intro}

One of the most intriguing results obtained so far at the Large Hadron Collider (LHC) is the indication of the lepton flavor universality violation (LFUV). 
First, from the measured partial branching fractions of $B\to K\ell^+\ell^-$, in the window of $q^2 \in [1,6] \ \gev^2$, the LHCb Collaboration in Ref.~\cite{Aaij:2014ora} reported
\begin{align}\label{exp:RK}
R_K = \frac{ \cb( B \to K \mu \mu)_{q^2\in [1,6]\gev^2}}{\cb( B \to K e e)_{q^2\in [1,6]\gev^2}} = 0.745 \pm^{0.090}_{0.074} \pm 0.036 \,,
\end{align}
which appears to be $2.4 \sigma$ below the Standard Model (SM) prediction, $R_K^{\rm SM} =1.00(1)$~\cite{Hiller:2003js}. 
Not many New Physics (NP) models can explain $R_K^{\rm exp} < R_K^{\rm SM}$, yet many attempts have been reported in the literature~\cite{RKpapers,Alonso:2015sja}. In terms of a generic low energy effective field theory it was soon realized that the models in which the NP contributions modify the couplings to muons, rather than to electrons, are more plausible. Furthermore it was understood that a modification of the couplings (Wilson coefficients) of muons to the scalar and/or pseudoscalar operator cannot generate the observed suppression, whereas a shift in couplings to the vector and/or axial operator can. Among those latter scenarios the popular are those that give rise to $C_9=-C_{10}$, or $C_9^\prime =-C_{10}^\prime$, patterns that are explicitly verified in several models, including those with an extra $Z^\prime$-boson, as well as the models which postulate the existence of low energy leptoquark states.

The hint that the loop induced decays $b\to s\ell\ell$ can break lepton flavor universality~\eqref{exp:RK} was corroborated by the most recent LHCb results~\cite{Bifani},
\begin{align}\label{exp:RKstar}
&R_{K^\ast}^\mathrm{low} = \frac{ \cb( B \to K^\ast \mu \mu)_{q^2\in [0.045,1.1]\gev^2}}{\cb( B \to K^\ast e e)_{q^2\in [0.045,1.1]\gev^2}} = 0.660 \pm^{0.110}_{0.070} \pm 0.024 \,,\nn\\[2ex]
&R_{K^\ast}^\mathrm{central} = \frac{ \cb( B \to K^\ast \mu \mu)_{q^2\in [1.1,6]\gev^2}}{\cb( B \to K^\ast e e)_{q^2\in [1.1,6]\gev^2}} = 0.685 \pm^{0.113}_{0.069} \pm 0.047 \,,
\end{align}
thus again $\sim (2.2-2.4) \sigma$ below the Standard Model (SM) prediction~\cite{Hiller:2003js}. If confirmed, these results would exclude the model of Ref.~\cite{Becirevic:2016yqi}, for example, in which the explanation of $R_K^{\rm exp} < R_K^{\rm SM}$ was made by means of a scalar leptoquark with hypercharge $Y=1/6$. That latter model verifies the pattern $\left( C_9^{\mu\mu}\right)^\prime  =-\left( C_{10}^{\mu\mu}\right)^\prime $, which entails $R_{K} <R_{K}^{\rm SM}$ and $\Rkst > R_{K^\ast}^\mathrm{SM}$. 

In this paper we will argue that another model with a low energy scalar leptoquark state can be used to explain both $R_K^{\rm exp} < R_K^{\rm SM}$ and 
$R_{K^\ast}^{\rm exp} <R_{K^\ast}^{\rm SM}$. In that model, also known as $R_2$-model, the leptoquark state transforms as $(3,2,7/6)$ under the Standard Model gauge group, $SU(3)_c \times SU(2)_L\times U(1)_Y$. A peculiarity of our model is that the coupling of leptoquark to $s$ and $\mu$ is absent and therefore the shift in $C_9^{\mu\mu}$ can be only achieved through loops. The model verifies  $C_9^{\mu\mu}=-C_{10}^{\mu\mu}$, and therefore both $R_K$ and $\Rkst$ can be smaller than in the Standard Model.  

The idea of explaining $R_K^{\rm exp}< R_K^{\rm SM}$ as a loop effect in a model with a scalar leptoquark is not new. In Ref.~\cite{Bauer:2015knc} the authors organized the Yukawa couplings in a similar way but in a model in which the scalar leptoquark is a weak singlet with hypercharge $Y=1/3$. It appeared that the dominant contribution, arising from the top-quark propagating in the loop, resulted in a positive shift of $C_9^{\mu\mu}$ so that the authors were obliged to compensate that effect with a very large charm-muon Yukawa coupling to comply with the general finding that $C_9^{\mu\mu}<0$. That induced problems elsewhere in phenomenology, ultimately making that particular model phenomenologically unviable, see discussion in Ref.~\cite{Becirevic:2016oho}. In our model, the dominant top quark contribution provides $C_9^{\mu\mu} < 0$, as needed, without inducing phenomenological problems elsewhere. Notice, however, that this model cannot explain $R_{D^{(\ast )}}$, the fact that was already (implicitly) shown in Ref.~\cite{Dorsner:2013tla}.

In the remainder of this paper we will describe the specifics of our model, compute the Wilson coefficients and describe the main constraints used to limit the values of the Yukawa couplings in Sec.~\ref{sec:eff}. We then discuss the phenomenology of the model in Sec.~\ref{sec:pheno} where we show that the model indeed accommodates $R_{K}^{\rm exp} <R_{K}^{\rm SM}$ entails $\Rkst^{\rm exp} <R_{K^\ast}^\mathrm{SM}$, consistent with hints from experiments. Other predictions are also made, in particular the rates of the lepton flavor violation modes. We conclude in Sec.~\ref{sec:conc}.

%%%%%%%%%%% 
%%%%%%%%%%%
\section{$\Delta^{(7/6)}$ or $R_2$ Model}
\label{sec:eff}

In this section we give the specifics of our particular model. To be able to relate it to the standard nomenclature, we first remind the reader of the low-energy effective theory
for $b\to s\ell\ell$ transitions, so that we can relate the results of our model to the relevant Wilson coefficients. 
 
 \subsection{Effective Hamiltonian}
 
Since we will also be interested in lepton flavor violation (LFV), we give the most general Hamiltonian describing the LFV transitions $b\to s\ell_1^- \ell_2^{+}$, with $\ell_{1,2}\in\{ e,\mu,\tau\}$, namely~\cite{Altmannshofer:2008dz},
\begin{equation}
\label{eq:hamiltonian}
\begin{split}
  \mathcal{H}_{\mathrm{eff}} = -\frac{4
    G_F}{\sqrt{2}}V_{tb}V_{ts}^* &\Bigg{\lbrace} \sum_{i=1}^6
  C_i(\mu)\mathcal{O}_i(\mu)+\sum_{i=7,8}
  \Big{[}C_i(\mu)\mathcal{O}_i(\mu)+\left(C_{i}(\mu)\right)^\prime \left(\mathcal{O}_{i}(\mu)\right)^\prime\Big{]}\\
& + \sum_{i=9,10,S,P}
  \Big{[} C^{\ell_1 \ell_2}_i(\mu)\mathcal{O}^{\ell_1 \ell_2}_i(\mu) + \left(C^{\ell_1 \ell_2}_{i}(\mu)\right)^\prime \left(\mathcal{O}^{\ell_1 \ell_2}_{i}(\mu)\right)^\prime\Big{]}\Bigg{\rbrace}
+\mathrm{h.c.},
\end{split}
\end{equation}
where the Wilson coefficients, $C_i(\mu)$ and $C_i^{\ell_1 \ell_2}(\mu)$, are associated with the following effective operators:
\begin{align}
\label{eq:C_LFV}
\begin{split}
\mathcal{O}_{9}^{\ell_1\ell_2}
  &=\frac{e^2}{(4\pi)^2}(\bar{s}\gamma_\mu P_{L}
    b)(\bar{\ell}_1\gamma^\mu\ell_2), \qquad\qquad\hspace*{0.4cm}
\mathcal{O}_{S}^{\ell_1\ell_2} =
  \frac{e^2}{(4\pi)^2}(\bar{s} P_{R} b)(\bar{\ell}_1 \ell_2),\\
    \mathcal{O}_{10}^{\ell_1\ell_2} &=
    \frac{e^2}{(4\pi)^2}(\bar{s}\gamma_\mu P_{L}
    b)(\bar{\ell}_1\gamma^\mu\gamma^5\ell_2),\qquad\qquad
\mathcal{O}_{P}^{\ell_1\ell_2} =
  \frac{e^2}{(4\pi)^2}(\bar{s} P_{R} b)(\bar{\ell}_1 \gamma^5 \ell_2),
\end{split}
\end{align}
in addition to the electromagnetic penguin operator,
$\mathcal{O}_7=e/(4\pi)^2m_b (\bar{s}\sigma_{\mu\nu}P_R b)F^{\mu\nu}$. The chirality flipped operators, $\mathcal{O}_i^\prime$, are obtained from the ones listed in Eq.~\eqref{eq:C_LFV}, after replacing $P_L\leftrightarrow P_R$. Using the above Hamiltonian, one can then compute the decay rates for $B_s\to\ell_1^-\ell^{+}_2$, $B\to K^{(\ast)}\ell_1^-\ell^{+}_2$, and other similar decay modes~\cite{Becirevic:2016zri,Gratrex:2015hna}.
As we mentioned in introduction, to obtain both $R_K < R_K^{\rm SM}$ and 
$R_{K^\ast} <R_{K^\ast}^{\rm SM}$ we need a NP contribution to the vector and axial Wilson coefficients, and in particular those coupling to the left-handed quark effective current, i.e. we need $C_{9,10}^{\mu\mu}\neq 0$.

%%%%%%%%%%%
%%%%%%%%%%% 
%%%%%%%%%%%
\subsection{$C_{9,10}^{\ell_1\ell_2}(\mu)$ in our $R_2$ Leptoquark Model}
\label{sec:models}

Leptoquarks are colored states mediating interactions between quarks and leptons. For a recent review of their properties see Ref.~\cite{Dorsner:2016wpm}. 
In general, a leptoquark can be a scalar or a vector field and it may come as a $SU(2)_L$-singlet, -doublet or -triplet~\cite{Buchmuller:1986zs}. 
Here we focus on the so-called $R_2$ model which involves a doublet of scalar leptoquarks with hypercharge $Y=7/6$. The general Yukawa Lagrangian for this model reads 
\begin{align}
\label{eq:slq1}
\begin{split}
\mathcal{L}_{\Delta^{(7/6)}} &= (g_R)_{ij} \bar{Q}_i{\boldsymbol\Delta}^{(7/6)}\ell_{Rj}+
                                                           (g_L)_{ij} \bar{u}_{Ri} { \widetilde{\boldsymbol\Delta}}^{(7/6) \dagger} L_{j}+ \mathrm{h.c.},\\[2.ex]
&=(V g_R)_{ij}\bar{u}_i P_R \ell_j\,\Delta^{(5/3)}+(g_R)_{ij}\bar{d}_i P_R \ell_j\,\Delta^{(2/3)} \\[1.3ex]
&\quad + (U g_L)_{ij}\bar{u}_i P_L \nu_j\,\Delta^{(2/3)}-(g_L)_{ij}\bar{u}_i P_L \ell_j\,\Delta^{(5/3)} +\mathrm{h.c.},
\end{split}
\end{align}
where $g_{L,R}$ are the matrices of Yukawa couplings, that we take to be
%\[
\bea
\label{eq:YC}
g_{L} = \left( \begin{matrix}
  0 & 0 & 0\\
  0 & g_{L}^{c \mu} & g_{L}^{c \tau}\\
  0 & g_{L}^{t \mu} & g_{L}^{t \tau}
\end{matrix}\right), \qquad g_{R} = \left( \begin{matrix}
  0 & 0 & 0\\
  0 & 0 &0\\
  0 & 0 & g_{R}^{b \tau}
\end{matrix}\right), \qquad V g_{R} = \left( \begin{matrix}
  0 & 0 &  V_{ub} g_{R}^{b \tau}\\
  0 & 0 & V_{cb} g_{R}^{b \tau}\\
  0 & 0 & V_{tb} g_{R}^{b \tau}
\end{matrix}\right), 
\eea
%\]
which is the main peculiarity of our model.~\footnote{As we shall see, after imposing the relevant constraints from the experimental data, $g_{R}^{b \tau}\approx 0$. The Yukawa couplings which verify that the matrices $g_{L}\neq 0$ and $g_{R}=0$ can stem from an underlying flavor symmetry. For example, this can be achieved by an extra $U(1)$ which, however, would necessitate introducing a second Higgs doublet. Details of this realization are clearly beyond the scope of the present paper. At low energy scales, the only relevant information about our model is the one given in Eq.~\eqref{eq:YC}. } 
%This particular choice can also be viewed as a result of an underlying flavor symmetry. 
The superscript in $\Delta^{(5/3)}$ and $\Delta^{(2/3)}$ refer to the electric charge of the two mass degenerate leptoquark states, $Q=Y+T_3$, where $Y$ is the hypercharge 
and $T_3$ the third component of weak isospin. Moreover, in Eq.~\eqref{eq:slq1} we use $Q_i = [(V^\dagger u_L)_i\; d_{Li}]^T$ and
$L_i = [(U\nu_L)_{i}\; \ell_{Li}]^T$, to denote the quark and lepton doublets, in which $V$ and $U$ are the
Cabibbo-Kobayashi-Maskawa (CKM) and the Pontecorvo-Maki-Nakagawa-Sakata (PMNS) matrices, respectively. Finally, $u_L$,
$d_L$, $\ell_L$ are the fermion mass eigenstates, whereas $\nu_L$ stand for the massless neutrino flavor eigenstates.  

\begin{figure}[ht!]
\centering
\includegraphics[width=0.5\linewidth]{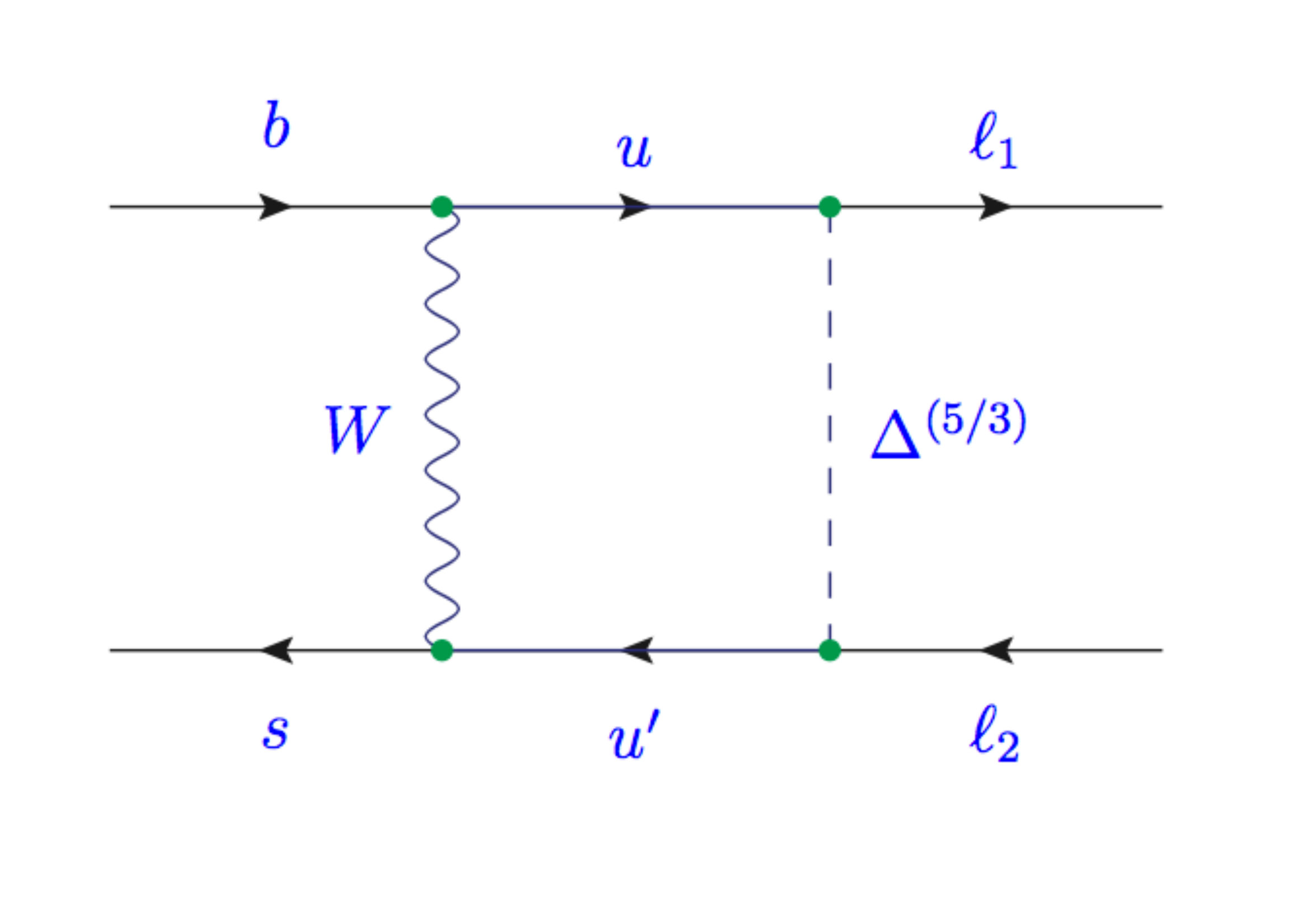}
\caption{\small \sl The only diagram contributing  $b\to s\ell_1\ell_2$ decay in the LQ scenario considered here. In a non-unitary gauge there is an extra diagram similar to 
the one depicted above, with $W$ replaced by a Goldstone boson. }
\label{fig:1}
\end{figure}

The above choice of Yukawa couplings, and in particular $g_R^{s\ell}=0$, means that the contributions of the leptoquark $\Delta^{(7/6)}$ to the transitions $b\to s\ell\ell$ can only be a loop effect and not 
generated at tree level as it is often the case in the scenario with low energy leptoquarks. The only diagram contributing (in the unitary gauge) is the one shown in Fig.~\ref{fig:1}.
%%%%%%%%%%%%%%%%%%%%%%%%%%%%%%%%%%%%%%%%
%%%%%%%%%%%%%%%%%%%%%%%%%%%%%%%%%%%%%%%%
%%%%%%%%%%%%%%%%%%%%%%%%%%%%%%%%%%%%%%%%
%%%%%%%%%%%%%%%%%%%%%%%%%%%%%%%%%%%%%%%%
We computed the corresponding amplitude, matched it onto the effective theory~\eqref{eq:hamiltonian}, and found
\begin{align}\label{eq:C9new}
C_9^{\ell_1\ell_2}=-C_{10}^{\ell_1\ell_2}= \sum_{u,u^\prime \in \{u,c,t\}} {V_{ub} V_{u^\prime s}^\ast\over V_{tb} V_{t s}^\ast } g_L^{u^\prime\ell_1} \left( g_L^{u \ell_2}\right)^\ast \mathcal{F}(x_u, x_{u^\prime})\,,
\end{align}
where $x_{i}= m_{i}^2/m_W^2$, and the loop function reads, 
\begin{align}
\mathcal{F}(x_u, x_{u^\prime})= {\sqrt{x_u x_{u^\prime}} \over 32\pi\alpha_\mathrm{em}}  &\biggl[ {  x_{u^\prime} (  x_{u^\prime} - 4) \log  x_{u^\prime}\over ( x_{u^\prime}-1) (x_u- x_{u^\prime})( x_{u^\prime}-x_\Delta)}
+ { x_u (  x_u - 4) \log  x_u\over ( x_u-1) (x_{u^\prime} - x_u )( x_u -x_\Delta)} 
\biggr.\nn\\[2.ex]
&\biggl.  - { x_\Delta (x_\Delta -4) \log x_\Delta \over (x_\Delta -1) (x_\Delta - x_u)( x_\Delta - x_{u^\prime} ) } 
\biggr]\,.
\end{align}
We checked that the above result is finite and gauge invariant by doing the computation in both the Feynman and the unitary gauge. The loop function vanishes when sending the quark mass to zero, and therefore the dominant contributions are those coming from $u=u^\prime=t$, and the one in which $u=t$, $u^\prime=c$, latter being CKM enhanced. 
This closes our discussion of the $R_2$ model with our particular setup specified by the structure of the $g_{L,R}$ matrices, as given in Eq.~\eqref{eq:YC}. 

\subsection{Constraints on $g_{L,R}^{\,q\ell}$}
\label{sec:constr}
The model described above can induce important contributions to some observables which have already been accurately measured. In other words, we check which quantity can be particularly sensitive to our model and then use its measured values to constrain the non-zero entries in the matrices $g_{L,R}$~\eqref{eq:YC}. 

First of all, by switching on the couplings to the leptoquark of the top quark and to $\mu$ and to $\tau$ leptons, one necessarily generates an extra term to the $\tau \to \mu\gamma$ decay amplitude. 
In order to comply with the experimentally established upper bound, $\mathcal{B}(\tau\to \mu \gamma)<4.4 \times 10^{-8}$ \cite{Aubert:2009ag}, we checked the expression derived in Ref.~\cite{Lavoura:2003xp,Dorsner:2016wpm} with which we agree, and write:
\begin{align}
\cb(\tau\to \mu \gamma)&=\ \tau_\tau\frac{\alpha_\mathrm{em} (m_\tau^2-m_\mu^2)^3}{4 m_\tau^3 } \left( |\sigma_L|^2 + |\sigma_R|^2 \right),\nn\\
\sigma_L &= 0\,, \nn\\
 \sigma_R &= {3 i m_\tau \over 64\pi^2 m_\Delta^2} \sum_{q\in \{c,t\}}  g_L^{q\mu \ast }  \biggl[ \   g_L^{q\tau} + \frac{2}{3} \frac{m_q}{m_\tau} V_{qb}g_R^{b\tau}\left( 1 + 4\log\frac{m_t^2}{m_\Delta^2} \right) \biggr] \,.
\end{align}
Since we need a significant value for $g_L^{t\mu}$ and $g_L^{c\mu}$ to describe the exclusive $b\to s\mu\mu$ decay rates, the above condition proves to be a severe bound on $g_R^{b\tau}$, due to the $m_t/m_\tau$ enhancement.

Another important constraint comes from the contributions to the muon's $g-2$. Current deviation between the measured and the SM values is $\Delta a_\mu^\mathrm{exp} = a_\mu^\mathrm{exp}- a_\mu^\mathrm{SM}=(2.8\pm 0.9)\times 10^{-9}$, where $a_\mu = (g_\mu -2)/2$, as usual. Since the SM estimate of this quantity is not yet fully assessed~\cite{nyffeler}, we require the leptoquark contribution to 
be smaller than $2\sigma$ error on $\Delta a_\mu^\mathrm{exp}$. To do so we use the expression~\cite{Dorsner:2016wpm}: 
\begin{align}
&\Delta a_\mu=-{3m_\mu^2\over 8\pi^2m_\Delta^2} \sum_{q\in \{c,t\}} |g_L^{q\mu}|^2 \left[ \frac{5}{3} f_S(m_q^2/m_\Delta^2) - f_F(m_q^2/m_\Delta^2) \right]\ ,\nn\\
&f_S(x) =\frac{x+1}{4 (1-x)^2}+\frac{x\log x}{2 (1-x)^3}\,,\qquad f_F(x) =\frac{x^2-5 x - 2}{12 (x-1)^3}+\frac{x\log x}{2 (1-x)^4}\,.
\end{align}

A very efficient constraint on $g_{L}^{\,t\ell}$ and $g_{L}^{\,c\ell}$ comes from the branching fractions $\cb(Z\to \ell\ell)$, which have been very accurately measured at LEP~\cite{Olive:2016xmw}: 
\bea\label{Zll}
\cb(Z\to \mu\mu)^\mathrm{exp}= 3.366(7)\,\%,\qquad \cb(Z\to \tau\tau)^\mathrm{exp}= 3.370(8)\,\% \,.
\eea
%%%%%%%%%%%%%%%%%%%%%%%%%%%%%%%%%%%%%%%%
%%%%%%%%%%%%%%%%%%%%%%%%%%%%%%%%%%%%%%%%
\begin{figure}[t!]
\centering
\includegraphics[width=0.25\linewidth]{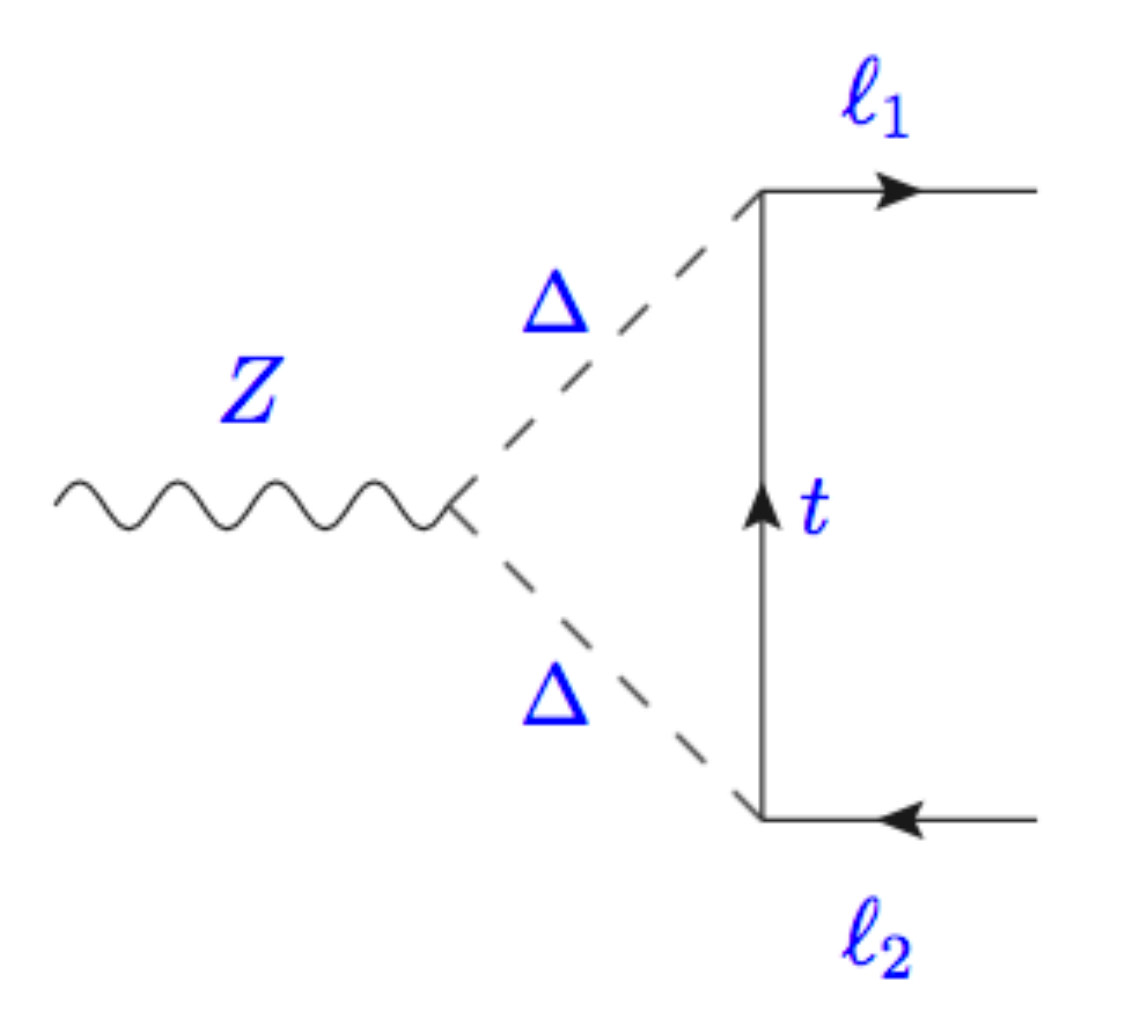}~\includegraphics[width=0.25\linewidth]{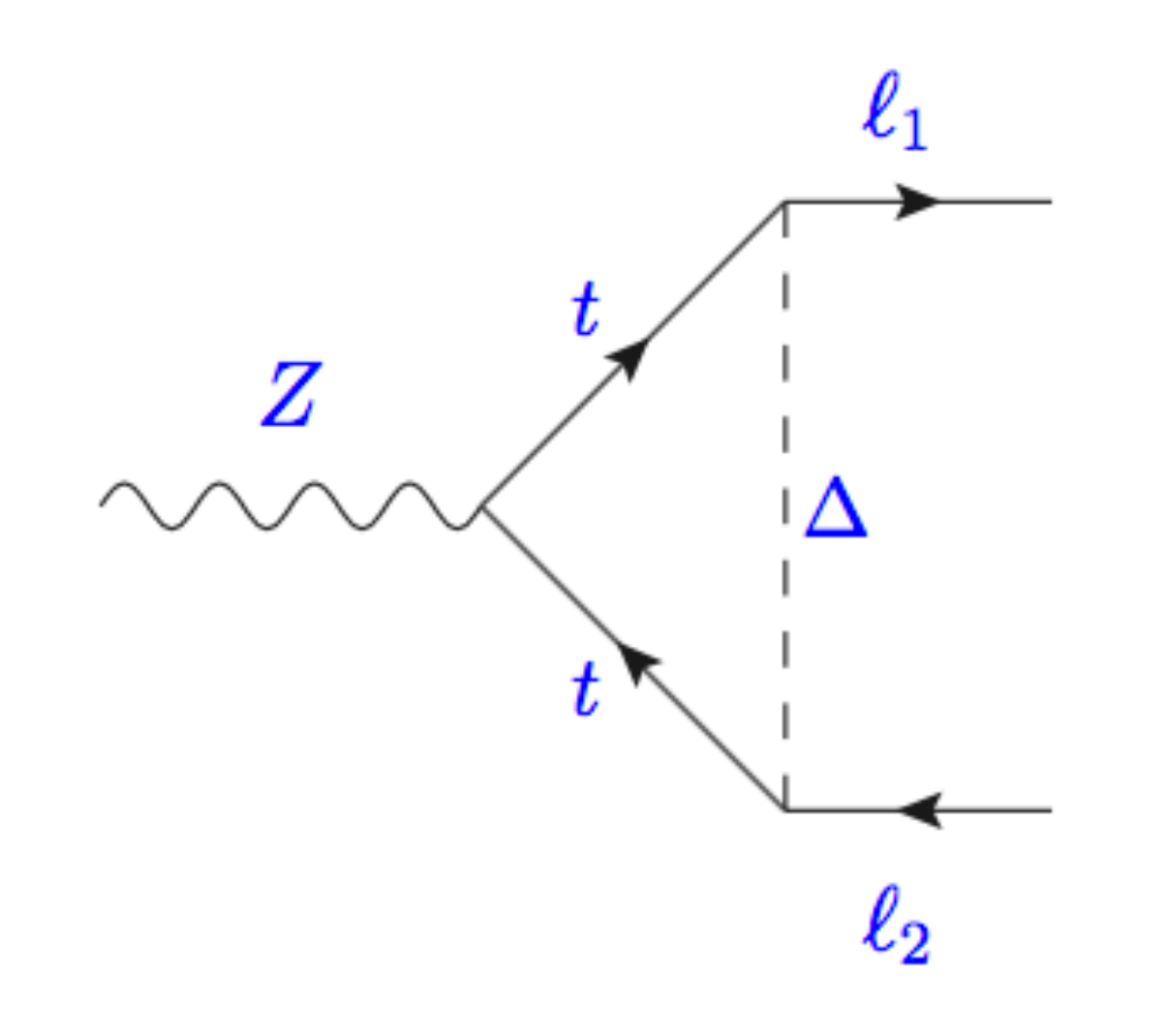}~\includegraphics[width=0.25\linewidth]{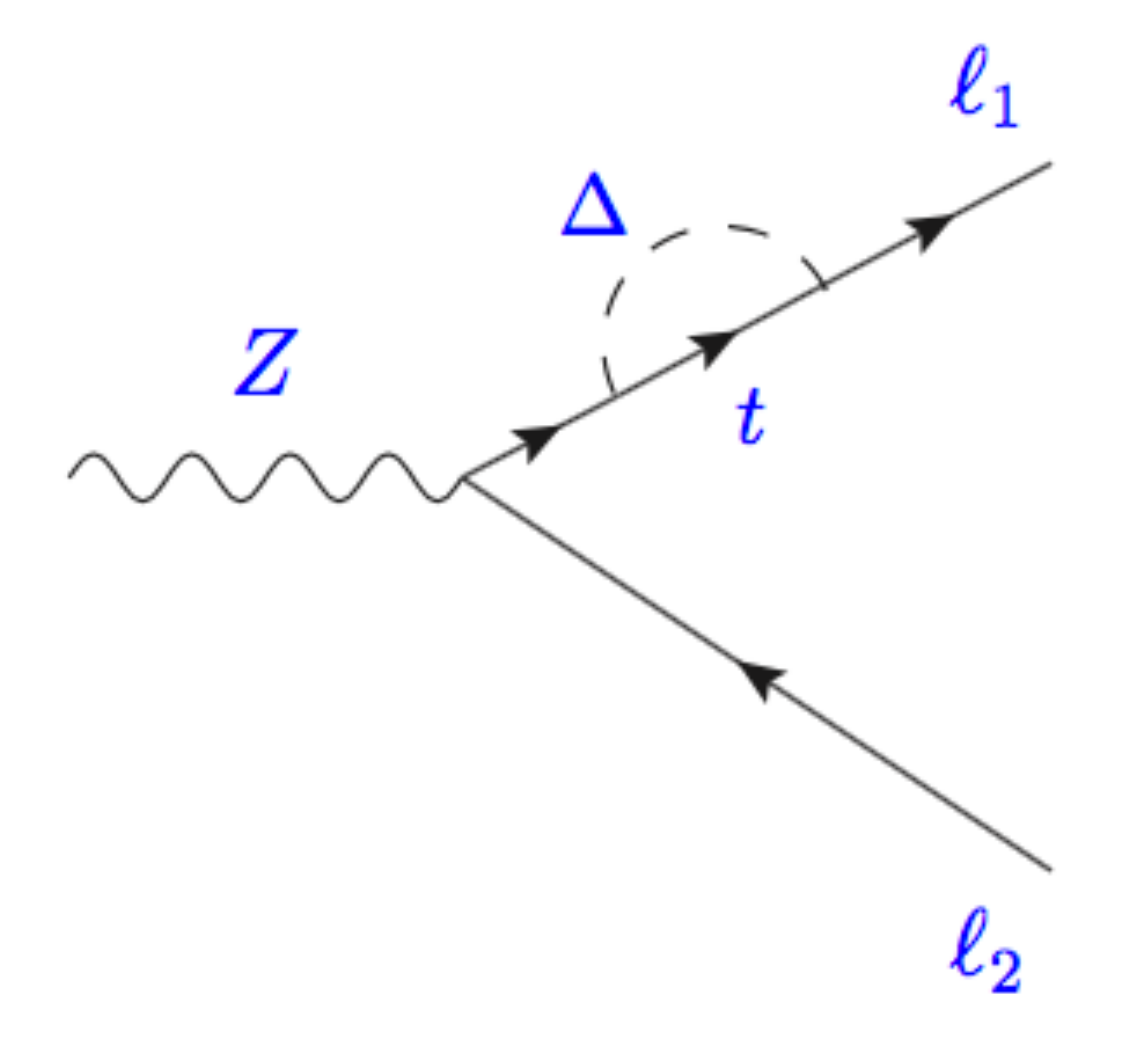}~\includegraphics[width=0.25\linewidth]{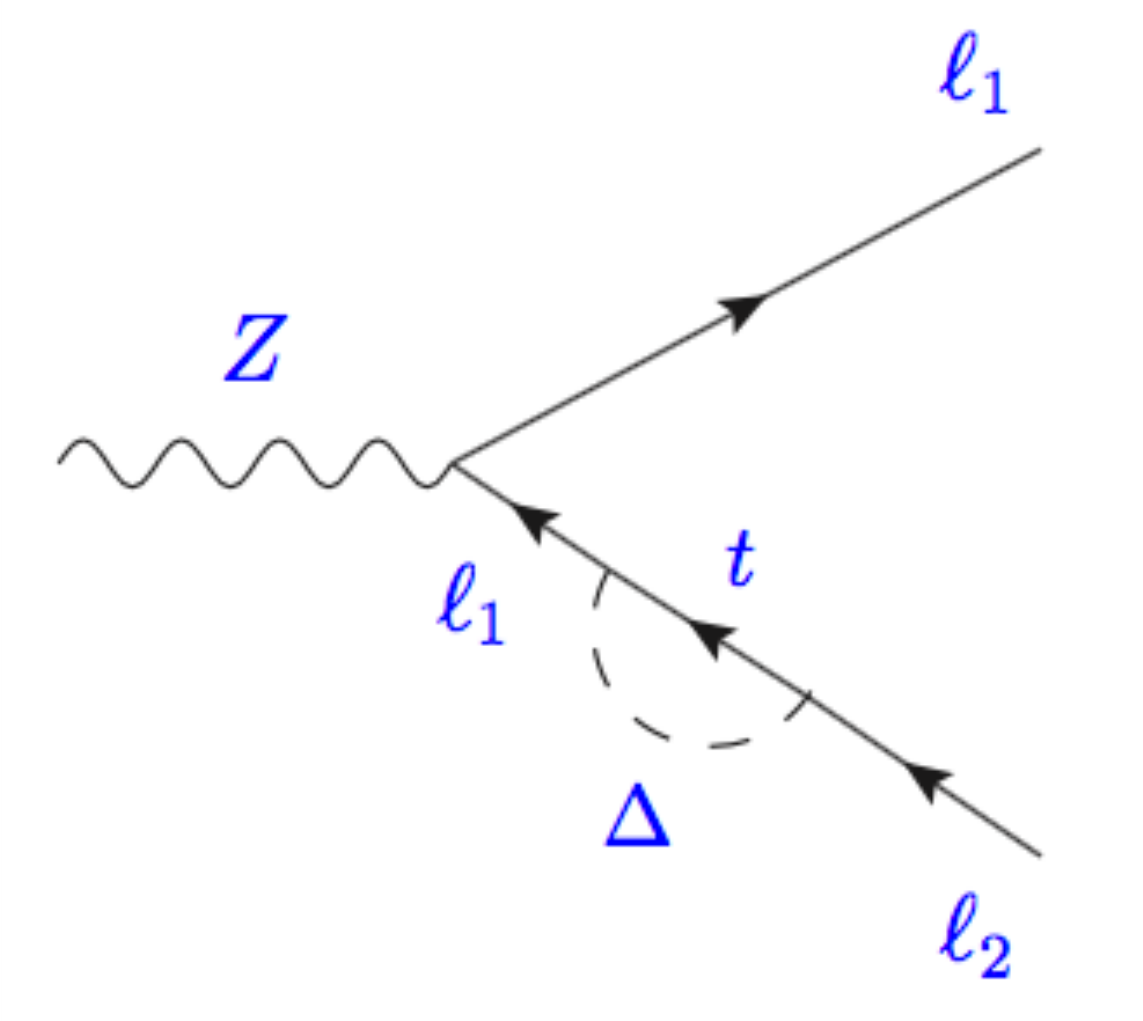}
\caption{\small \sl Contributions to the $Z\to \ell_1\ell_2$ decay amplitude generated in our $R_2$ model, where $\Delta \equiv \Delta^{(5/3)}$. Another set of diagrams, similar to those shown above is obtained by replacing $t\to c$. }
\label{fig:2}
\end{figure}
%%%%%%%%%%%%%%%%%%%%%%%%%%%%%%%%%%%%%%%%
%%%%%%%%%%%%%%%%%%%%%%%%%%%%%%%%%%%%%%%%
In our model the diagrams contributing to $Z\to \ell\ell$ (or, more generally, to $Z\to \ell_1\ell_2$) are shown in Fig.~\ref{fig:2}.
We computed the full amplitude, matched it with the effective Lagrangian,
\bea\label{eq:LZ}
\mathcal{L}_\mathrm{eff} = \frac{g}{2\cos\theta_W} C_{VL}^{\ell_1\ell_2}\, \bar\ell_1 \gamma^\mu P_L \ell_2 Z^\mu\, ,
\eea 
and obtained, 
\begin{align}\label{eq:CVL}
C_{VL}^{\ell_1\ell_2} = - {3\over 16\pi^2 }
\Bigg\{ & g_L^{t \ell_1} g_L^{t \ell_2 \ast}\, \frac{m_t^2}{m_\Delta^2}  \left( 1 + \log\frac{m_t^2}{m_\Delta^2} \right) \nn\\
- & \frac{4}{9}  g_L^{c \ell_1} \, g_L^{c \ell_2 \ast}  \frac{m_Z^2}{m_\Delta^2} 
 \left[ 
 \sin^2\theta_W  \left(  \log\frac{m_Z^2}{m_\Delta^2} + i\pi +\frac{1}{12} 
 \right) - \frac{1}{8} \right] 
\Bigg\}.
\end{align}
The top contribution in the above formula agrees with the result of Ref.~\cite{ColuccioLeskow:2016dox} while the contribution arising from charm is new. 
Using the Lagrangian~\eqref{eq:LZ}, we then obtain  
\bea
\cb(Z\to \ell\ell)={ m_Z^3\over 24\pi v^2 \Gamma_Z}\biggl[ \vert C_{VL}^{\ell\ell}\vert^2 - 2 \, \mathrm{Re}(1+ C_{VL}^{\ell\ell}) \cos(2\theta_W) + 2 + \cos(4\theta_W)\biggr]\,.
\eea
In practice, we find it more convenient to consider 
\bea
R_Z^{\ell\ell} = {\cb(Z\to \ell\ell)\over \cb(Z\to \ell\ell)^\mathrm{SM} }\,,
\eea
and to use the values~(\ref{Zll}) to $2\sigma$ accuracy.

Finally, the major constraint on the model comes from the exclusive $b\to s\mu\mu$ decays. Like in our previous publications, we prefer to use two 
most reliable decay modes, as far as hadronic uncertainties are concerned, namely $B_s\to \mu\mu$ and $B\to K\mu\mu$. More specifically, to 
compare the measured $\mathcal{B}(B_s\to\mu^+\mu^-)^{\mathrm{exp}}=(3.0\pm 0.6^{+0.3}_{-0.2})\times
10^{-9}$~\cite{CMS:2014xfa} with theory, the only needed quantity is the decay constant $f_{B_s}$ which has been computed by many lattice QCD collaborations. 
The most recent average value is $f_{B_s}=224(5)$~MeV~\cite{Aoki:2016frl}. Similarly, the lattice QCD results for the $B\to K$ form factors have been computed 
at large values of $q^2$  by two collaborations~\cite{Bouchard:2013pna,Bailey:2015dka}. Their results agree and can be used to compare with the measured 
$\mathcal{B}(B\to K \mu^+\mu^-)_{q^2 \in [15,22]\gev^2}^{\mathrm{exp}}=(8.5\pm 0.3\pm 0.4 )\times10^{-8}$~\cite{Aaij:2014pli}. 
Adding to the SM values of the Wilson coefficients the New Physics ones, which in our model means $C_{9,10}^{\mu\mu}$ satisfying the condition $C_9^{\mu\mu}=-C_{10}^{\mu\mu}$, to $2\sigma$ accuracy, from the fit between the measured and theory values for $\mathcal{B}(B\to K \mu^+\mu^-)_{q^2\in  [15,22]\gev^2}$ and for $\mathcal{B}(B_s\to\mu^+\mu^-)$, one extracts~\cite{Becirevic:2016oho}
\begin{align}
\label{eq:c9c10}
C_{9}^{\mu\mu}=-C_{10}^{\mu\mu}\in (-0.76,-0.04)\, . 
\end{align}
This value together with the expression~\eqref{eq:C9new}, with $\ell_1=\ell_2=\mu$, leads to stringent bounds on the couplings we are interested in, and ultimately provides $R_{K,K^\ast}<1$, as we shall see below. The result of this procedure will be called ``Fit A".

Another possibility is to, in addition to the above two quantities, also consider a few ``clean" quantities extracted from the study of $B\to K^\ast \mu\mu$ decay mode. In particular, the measured $\mathcal{B}(B\to K^\ast \mu^+\mu^-)_{q^2\in [15,19]\gev^2}^\mathrm{exp}=1.95(16)\times 10^{-7}$~\cite{Aaij:2016flj} can be combined with form factors computed on the lattice at large values of $q^2$~\cite{Horgan:2013hoa}. Furthermore, the three observables obtained from the decay's angular distribution, all three depending only on the so-called transverse amplitudes, $A_{\parallel , \perp}(q^2)$, with respect to the spin of the on-shell $K^\ast$. These quantities, which also appear to be very mildly sensitive to hadronic uncertainties~\cite{Becirevic:2011bp}, are known as 
$A_T^{(2)}$, $ A_T^{({\rm re})}$ and $A_T^{({\rm im})}$, are translated into $P_{1,2,3}$ in Ref.~\cite{Matias:2012xw}, the notation also respected by the experimentalists.~\footnote{
Note that, $P_1\equiv A_T^{(2)}$, $P_2\equiv A_T^{({\rm re})}/2$, $P_3\equiv -A_T^{({\rm im})}/2$, where we take into account the correct signs~\cite{Gratrex:2015hna} to correctly compare with experimental results.} More specifically, 
\begin{align}\label{eq:Pi}
&P_1=  \frac{\left< |A_\perp^{L,R}|^2-|A_\parallel^{L,R}|^2\right>}{\left<  |A_\perp^{L,R}|^2+|A_\parallel^{L,R}|^2\right>},&\qquad P_1^\mathrm{exp} &= \{ 0.08(25)_{\mathrm{low}\, q^2}, -0.50(10)_{\mathrm{high}\, q^2}\},\nn\\
&P_2= - \frac{\left< \re\left[ A_\perp^{L} A_\parallel^{L\, \ast} - A_\perp^{R} A_\parallel^{R\, \ast}\right]\right>}{\left< |A_\perp^{L,R}|^2+|A_\parallel^{L,R}|^2\right>} , & P_2^\mathrm{exp}&=\{-0.16(7)_{\mathrm{low}\, q^2}, 0.36(3)_{\mathrm{high}\, q^2}\}, \nn\\
&P_3= \frac{\left< \im\left[ A_\perp^{L} A_\parallel^{L\, \ast} - A_\perp^{R} A_\parallel^{R\, \ast}\right]\right>}{\left< |A_\perp^{L,R}|^2+|A_\parallel^{L,R}|^2\right>} ,& P_3^\mathrm{exp}&=\{ 0.21(14)_{\mathrm{low}\, q^2}, 0.08(6)_{\mathrm{high}\, q^2}\} ,
\end{align}
where the full expressions for $A_{\parallel , \perp}\equiv A_{\parallel , \perp}(q^2)$, in terms of form factors and Wilson coefficients, can be found eg. in Ref.~\cite{Becirevic:2016zri}. In the above notation,  $\langle\dots \rangle$ means that the numerator and denominator have been partially integrated over a specific window of $q^2$. The experimental values for $P_{1,2,3}$ in two (wide) bins, corresponding to $q^2\in [1.1,6]\, \gev^2$ and $q^2\in [15,19]\, \gev^2$, which are referred to as ``low $q^2$" and ``high $q^2$", are extracted from Ref.~\cite{Aaij:2016flj}. 
Thus, from the fit in which we use 
\bea
&&\cb(B_s\to \mu\mu),\ \mathcal{B}(B\to K \mu\mu)_{q^2\in [15,22]\gev^2}, \mathcal{B}(B\to K^\ast \mu\mu)_{q^2\in [15,19]\gev^2}, \nn\\[2.ex]
&& (P_1,P_2,P_3)_{\mathrm{low}\, q^2}, \  (P_1,P_2,P_3)_{\mathrm{high}\, q^2} ,
\eea
to $2\sigma$ accuracy, we obtain 
\begin{align}
\label{eq:c9c10bis}
C_{9}^{\mu\mu}= -C_{10}^{\mu\mu}\in (-0.70,-0.16)\, . 
\end{align}
which will be referred to as ``Fit B".

Before closing this section we believe it is worth emphasizing that the model we consider here does not give any contribution to the $B_s-\overline B_s$ mixing amplitude (at the one-loop level).

\section{Phenomenology: $R_K$ and $R_{K^\ast}$}
\label{sec:pheno}

\begin{table}[ht!]
\centering
\renewcommand{\arraystretch}{1.75}
\begin{tabular}{|c|c|c|}
\hline 
Quantity & Fit A & Fit B  \\ \hline\hline
$R_K$ (low~$q^2$) & $[0.64, 0.96]$  &  $[0.66, 0.91]$   \\  
$\Rkst$ (low~$q^2$) & $[0.83, 0.92]$  &  $[0.84, 0.91]$   \\  \hline
$R_K$ (central~$q^2$) & $[0.66, 0.98]$  &  $[0.69, 0.93]$   \\  
$\Rkst$ (central~$q^2$) & $[0.67, 0.98]$  &  $[0.69, 0.93]$   \\  \hline
$R_K$ (high~$q^2$) & $[0.65, 0.98]$  &  $[0.68, 0.93]$   \\  
$\Rkst$ (high~$q^2$) & $[0.64, 0.98]$  &  $[0.67, 0.92]$   \\  \hline
\end{tabular}
\caption{\small \sl Intervals of $R_K$ and $\Rkst$ obtained solely from the values for the Wilson coefficient $C_9^{\mu\mu}= - C_{10}^{\mu\mu}$ obtained from the Fit A [Eq.~\eqref{eq:c9c10}] 
and Fit B [Eq.~\eqref{eq:c9c10bis}], as discussed in the text. 
}
\label{tab:1} 
\end{table}
%%%%%%%%%%%%%%%%%%%%%%%%%%%%%%%%%%%%%%%%
%%%%%%%%%%%%%%%%%%%%%%%%%%%%%%%%%%%%%%%%
\begin{figure}[t!]
\centering
\includegraphics[width=0.42\linewidth]{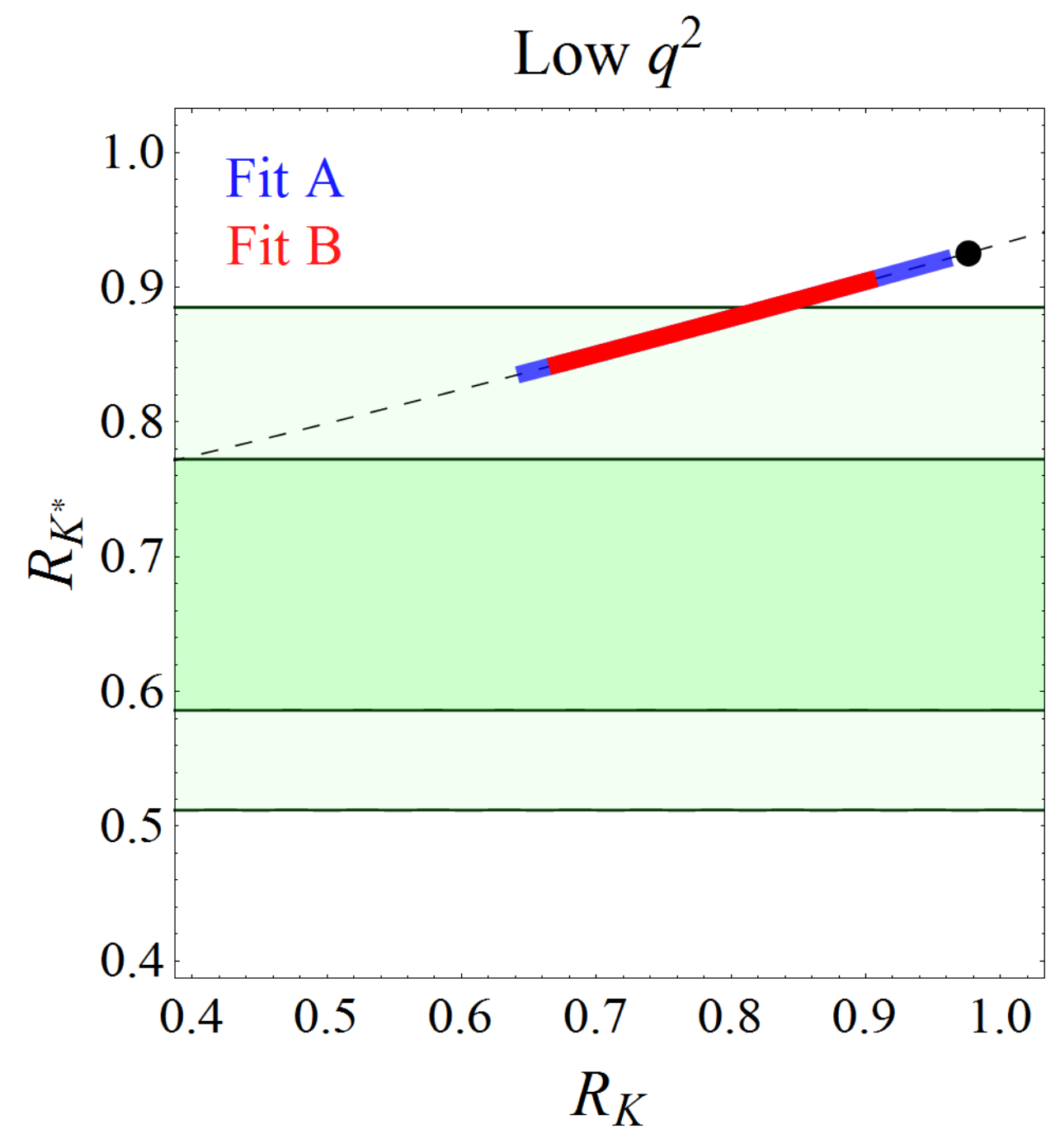}~\includegraphics[width=0.42\linewidth]{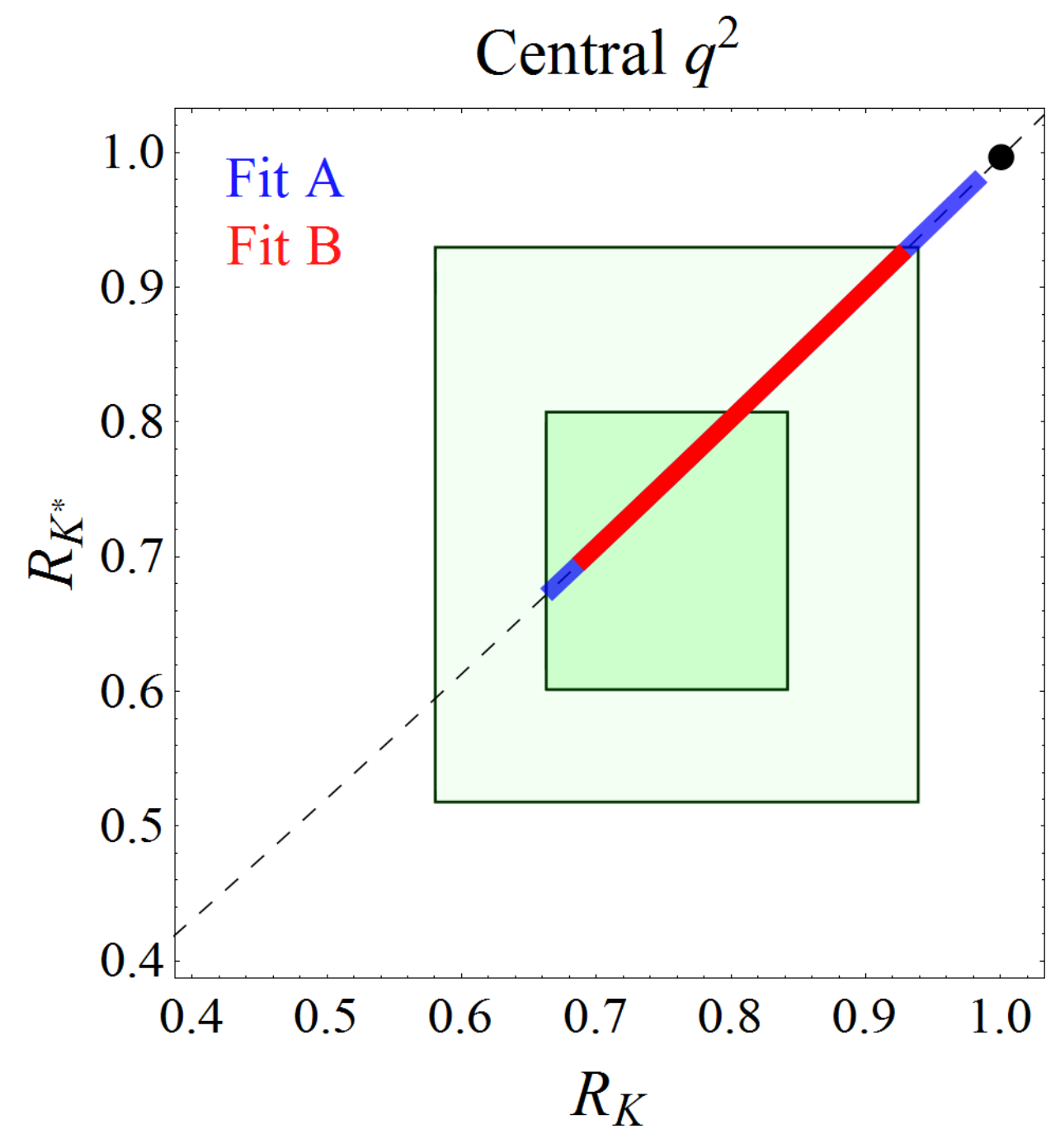}\\
\includegraphics[width=0.4\linewidth]{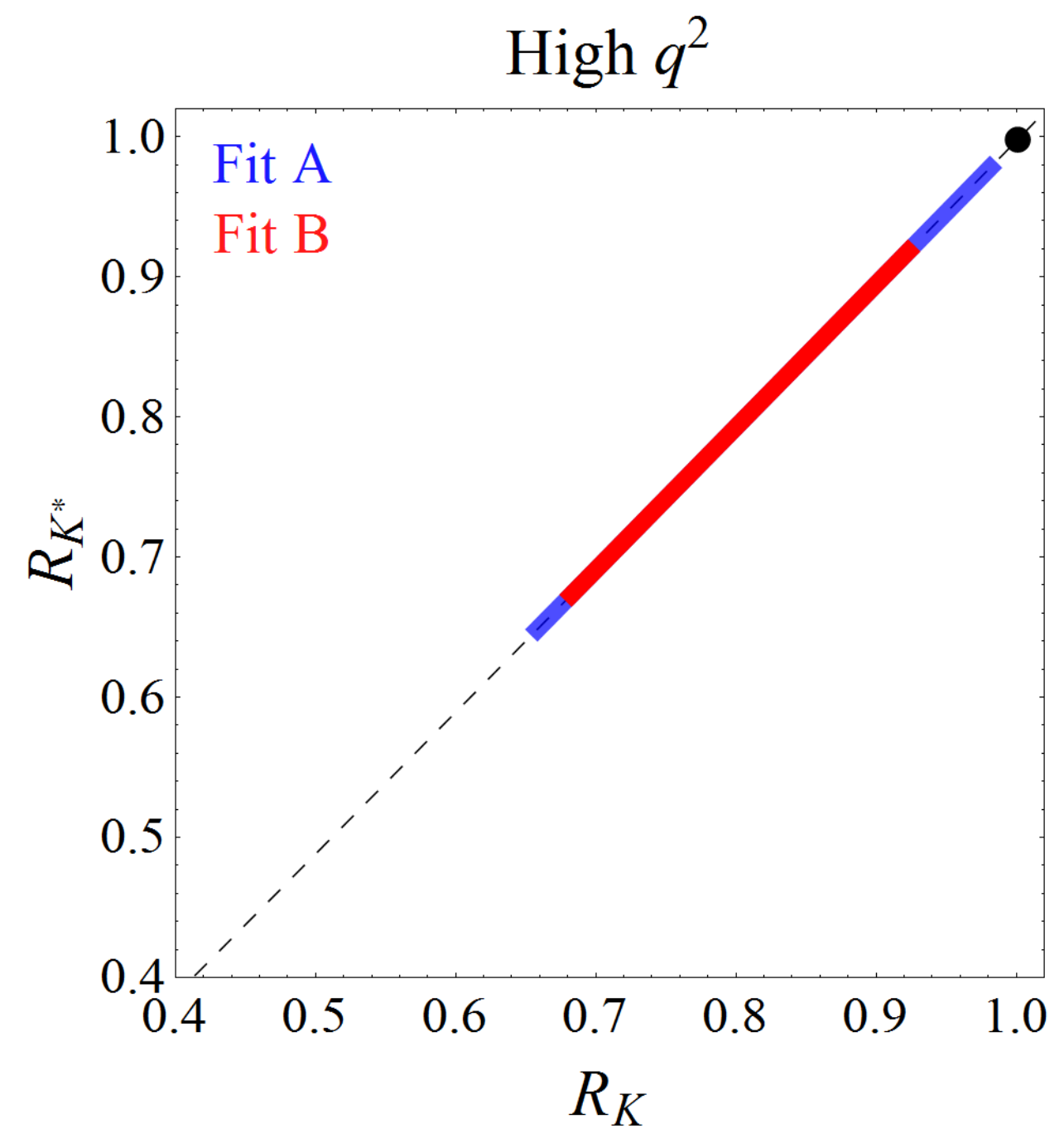}
\caption{\small \sl Results for $R_K$ and $\Rkst$ obtained solely from the values for the Wilson coefficient $C_9^{\mu\mu}= - C_{10}^{\mu\mu}$ obtained from the Fit A [Eq.~\eqref{eq:c9c10}] 
and Fit B [Eq.~\eqref{eq:c9c10bis}] as discussed in the text. The shaded area correspond to the measured values to $1\sigma$ and $2\sigma$, cf. Eqs.~(\ref{exp:RK},\ref{exp:RKstar}). The thick dot corresponds to the SM result.
}
\label{fig:3}
\end{figure}
%%%%%%%%%%%%%%%%%%%%%%%%%%%%%%%%%%%%%%%%
%%%%%%%%%%%%%%%%%%%%%%%%%%%%%%%%%%%%%%%%
We are now in a position to discuss the phenomenology of our model, including the main topic of this paper,  
$R_K< R_K^\mathrm{SM}$ and $R_{K^\ast}<R_{K^\ast}^\mathrm{SM}$, the problem addressed by a number of authors in Ref.~\cite{interpretations_day1}.

Before focusing on our model, we first use the results for $C_9^{\mu\mu}=-C_{10}^{\mu\mu}$ given in Eqs.~\eqref{eq:c9c10} and \eqref{eq:c9c10bis}, respectively referred to as Fit A and Fit B, and compute 
\bea
R_{K^{(\ast)}}=   \frac{ \cb( B \to K^{(\ast )} \mu \mu)_{q^2\in [q_1^2,q_2^2]} }{\cb( B \to K^{(\ast )} e e)_{q^2\in [q_1^2,q_2^2]} } \,,
\eea
by relying on the expressions given in our Refs.~\cite{Becirevic:2016oho,Becirevic:2016zri}, and for three separate intervals in $q^2$. To make the comparison with experiment easier we consider three intervals: 
$q^2\in [0.045,1.1]\,\gev^2$, $[1.1,6]\,\gev^2$ and $[15,19]\,\gev^2$ and call them low, central and large $q^2$-bin, respectively.

In Fig.~\ref{fig:3} we plot the resulting $R_K$ and $\Rkst$ by relying only on the effective theory and on the value of $C_9^{\mu\mu}=-C_{10}^{\mu\mu}$ obtained from the fit with the data as discussed above. We see that in the central bin our results are in very good agreement with experiment, at the $1\sigma$ level, 
regardless of the $C_9^{\mu\mu}$ value we use, \eqref{eq:c9c10} or \eqref{eq:c9c10bis}. 
The situation is not as favorable in the low $q^2$-bin, in which the agreement between ours and the measured values of $\Rkst$ is not better than $1.5\sigma$. This, however, is a very good agreement too. The values shown in Fig.~\ref{fig:3} are also listed in Tab.~\ref{tab:1}.
We stress again that these results are relevant to any scenario satisfying the pattern $C_9^{\mu\mu}= - C_{10}^{\mu\mu}$, and therefore our model in particular.

We next focus on our model and beside $C_9^{\mu\mu}$ obtained in Eqs.~\eqref{eq:c9c10} and~\eqref{eq:c9c10bis} we also use the constraints discussed in the previous section. These constraints appear to be quite severe. Consistency with $C_9^{\mu\mu}$ requires rather large values of the muonic couplings to the leptoquark. For that reason, the experimental bound on $\cb(\tau\to\mu\gamma)$ will necessarily restrain $g_R^{b\tau}$ to very small values. 
The values of $g_L^{t\mu}$ [$g_L^{t\tau}$] and $g_L^{c\mu}$ [$g_L^{c\tau}$] are then saturated by $\Delta a_\mu$ and by the required consistency with the measured $\cb (Z\to \mu\mu)$ [$\cb (Z\to \tau\tau)$].

\

\begin{figure}[t!]
\centering
\includegraphics[width=0.46\linewidth]{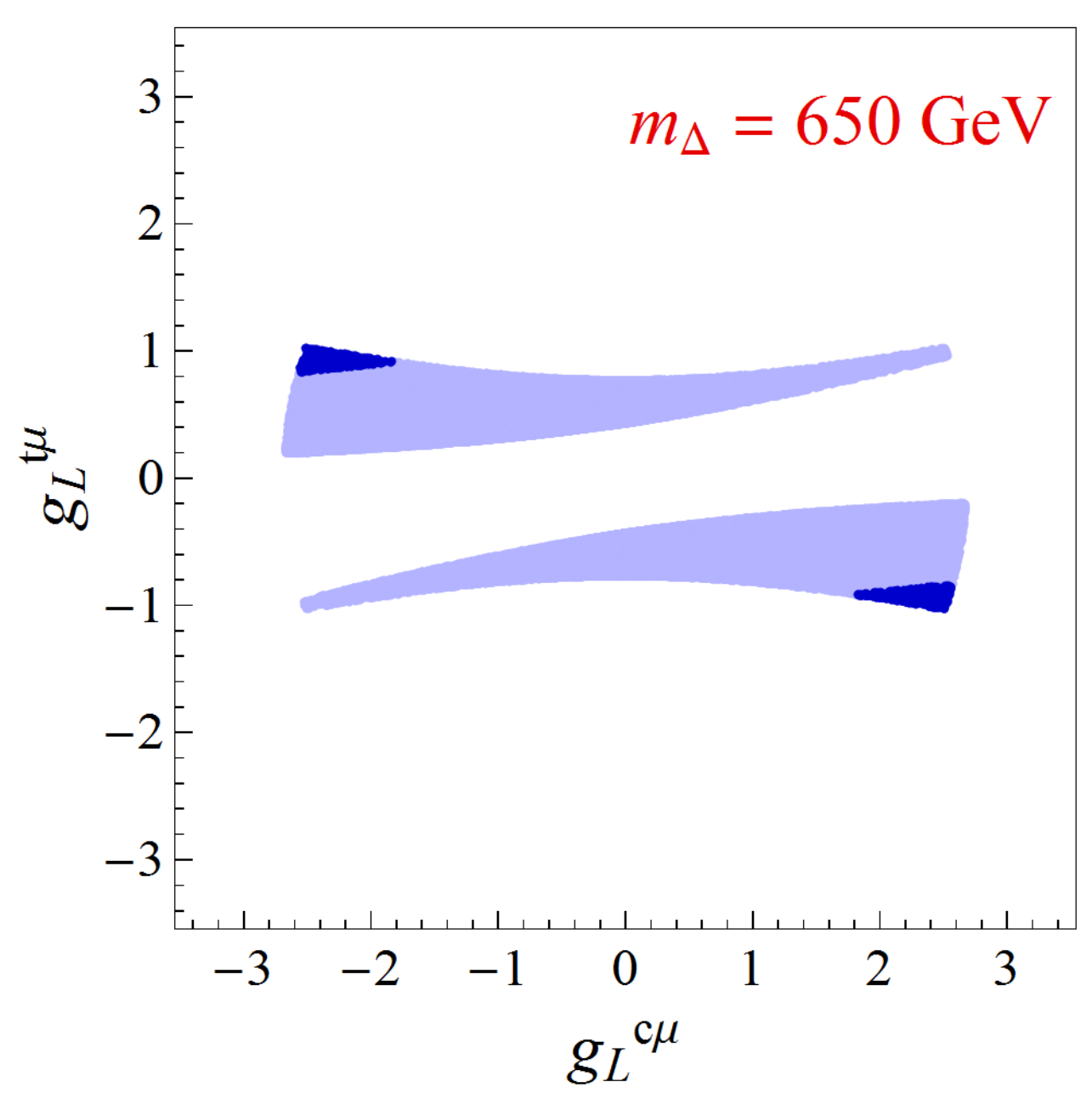}~\includegraphics[width=0.46\linewidth]{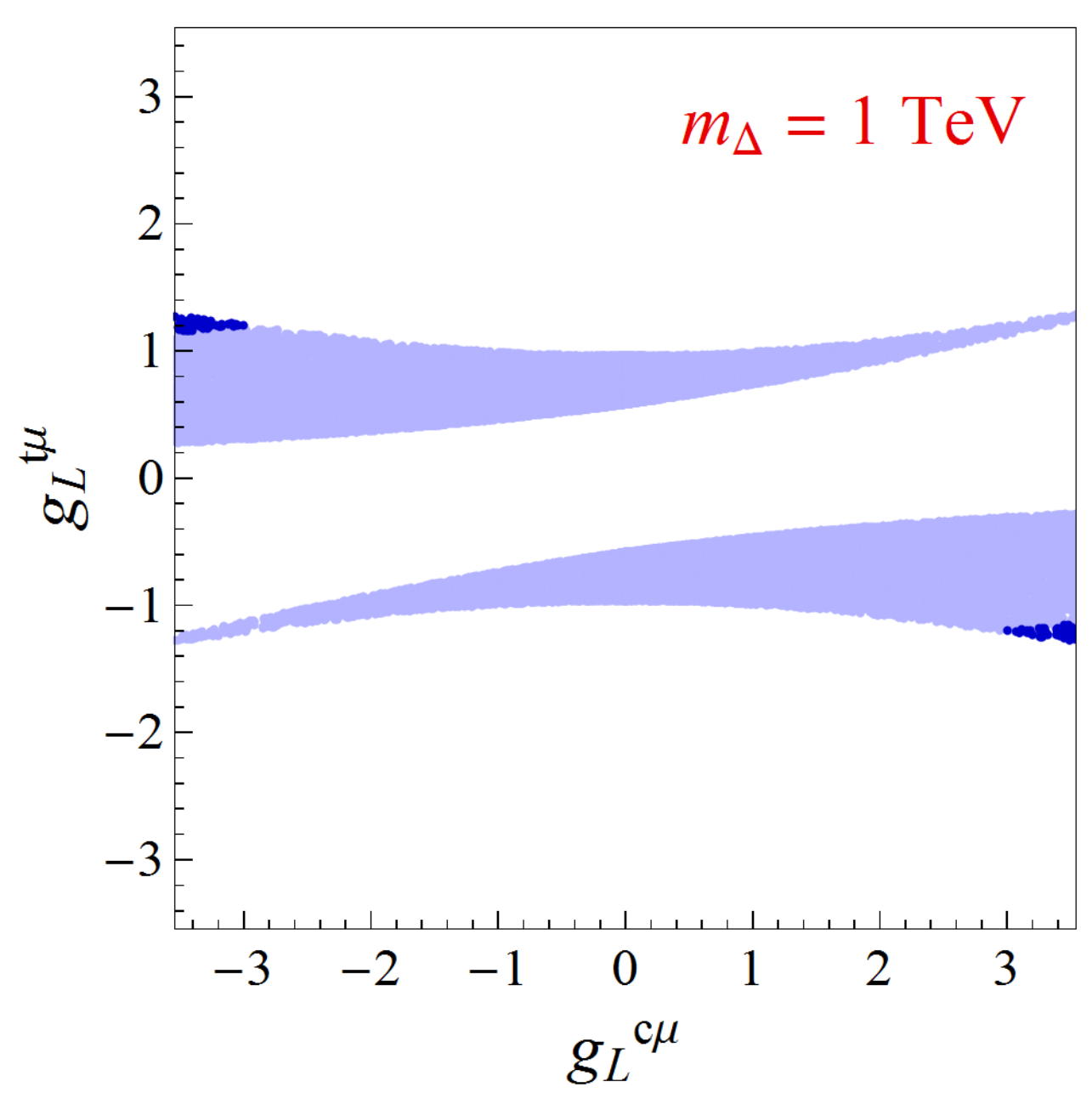}\\
\caption{\small \sl 
Allowed values for the couplings $g_L^{\ t\mu}$ and  $g_L^{\ c\mu}$ consistent with all the constraints discussed in the previous section. Plots are provided for $m_\Delta =650$~GeV and $m_\Delta =1$~TeV. Highlighted regions correspond to the values of the couplings that ensure the $1.5\sigma$ agreement of $R_{K}$ and $R_{K^\ast}$ with experiment in the central $q^2$-bin.
}
\label{fig:4}
\end{figure}

We performed several scans of the model parameters. We first fixed the mass of leptoquark to either $m_\Delta =650$~GeV or to $m_\Delta =1$~TeV, and varied all the couplings 
within $|g_{L,R}^{q\ell}|\leq \sqrt{4\pi}$. As we anticipated above, the allowed values of $g_R^{b\tau}$ are indeed negligibly small, and for our phenomenological purposes this coupling can be safely neglected. In the following we set it to zero. On the other hand, constraints on the couplings to muon result in the regions shown in Fig.~\ref{fig:4}. Clearly, for larger $m_\Delta$ the couplings grow and for reasonable values of $m_\Delta$ (less than a few TeV) the only coupling that hits the perturbativity bound is $g_L^{\,c\mu}$ while the other ones remain well bellow $\sqrt{4\pi}$. 
%%%%%%%%%%%%%%%%%%%%%%%%%%%%%%%%%%%%%%%%
%%%%%%%%%%%%%%%%%%%%%%%%%%%%%%%%%%%%%%%%
Notice also that in Fig.~\ref{fig:4} we highlight the regions of couplings that are needed to provide a $1.5\sigma$ compatibility of $R_K$ and $\Rkst$ with experimental results in the central $q^2$-bin. 
In other words, to get close to the measured values of $R_K^\mathrm{exp}$ and $R_{K^\ast}^\mathrm{exp}$ the values of couplings $g_L^{\,c\mu}$ and $g_L^{\,t\mu}$ indeed need to be large (larger than $1$). 
The values for $R_K$ and $R_{K^\ast}$ obtained with our model are given in Tab.~\ref{tab:2}. We see that the situation regarding the agreement with experimental values~\eqref{exp:RK} and \eqref{exp:RKstar} remains similar to the discussion based only on $C_9^{\mu\mu}$, i.e. our values for $R_K$ and $\Rkst$ are compatible with experiment in the central $q^2$-bin to $1.1\sigma$, while in the low-$q^2$-bin the agreement of our $\Rkst$ with the value in~\eqref{exp:RKstar} is at the $1.8\sigma$ level.
\begin{table}[ht!]
\centering
\renewcommand{\arraystretch}{1.75}
\begin{tabular}{|c|c|c|}
\hline 
Quantity & $m_\Delta=650$~GeV & $m_\Delta=1$ TeV  \\ \hline\hline
$R_K$ (low~$q^2$) & $[0.80, 0.96]$  &  $[0.82, 0.96]$   \\  
$\Rkst$ (low~$q^2$) & $[0.88, 0.92]$  &  $[0.88, 0.92]$   \\  \hline
$R_K$ (central~$q^2$) & $[0.82, 0.98]$  &  $[0.85, 0.98]$   \\  
$\Rkst$ (central~$q^2$) & $[0.82, 0.98]$  &  $[0.85, 0.98]$   \\  \hline
$R_K$ (high~$q^2$) & $[0.81, 0.98]$  &  $[0.84, 0.98]$   \\  
$\Rkst$ (high~$q^2$) & $[0.81, 0.98]$  &  $[0.83, 0.98]$   \\  \hline
\end{tabular}
\caption{\small \sl Intervals of $R_K$ and $\Rkst$ obtained in our model by using all the constraints discussed in the text, and  $C_9^{\mu\mu}= - C_{10}^{\mu\mu}$ in Eq.~\eqref{eq:c9c10} in particular.
}
\label{tab:2} 
\end{table}

\begin{figure}[ht!]
\centering
\includegraphics[width=0.46\linewidth]{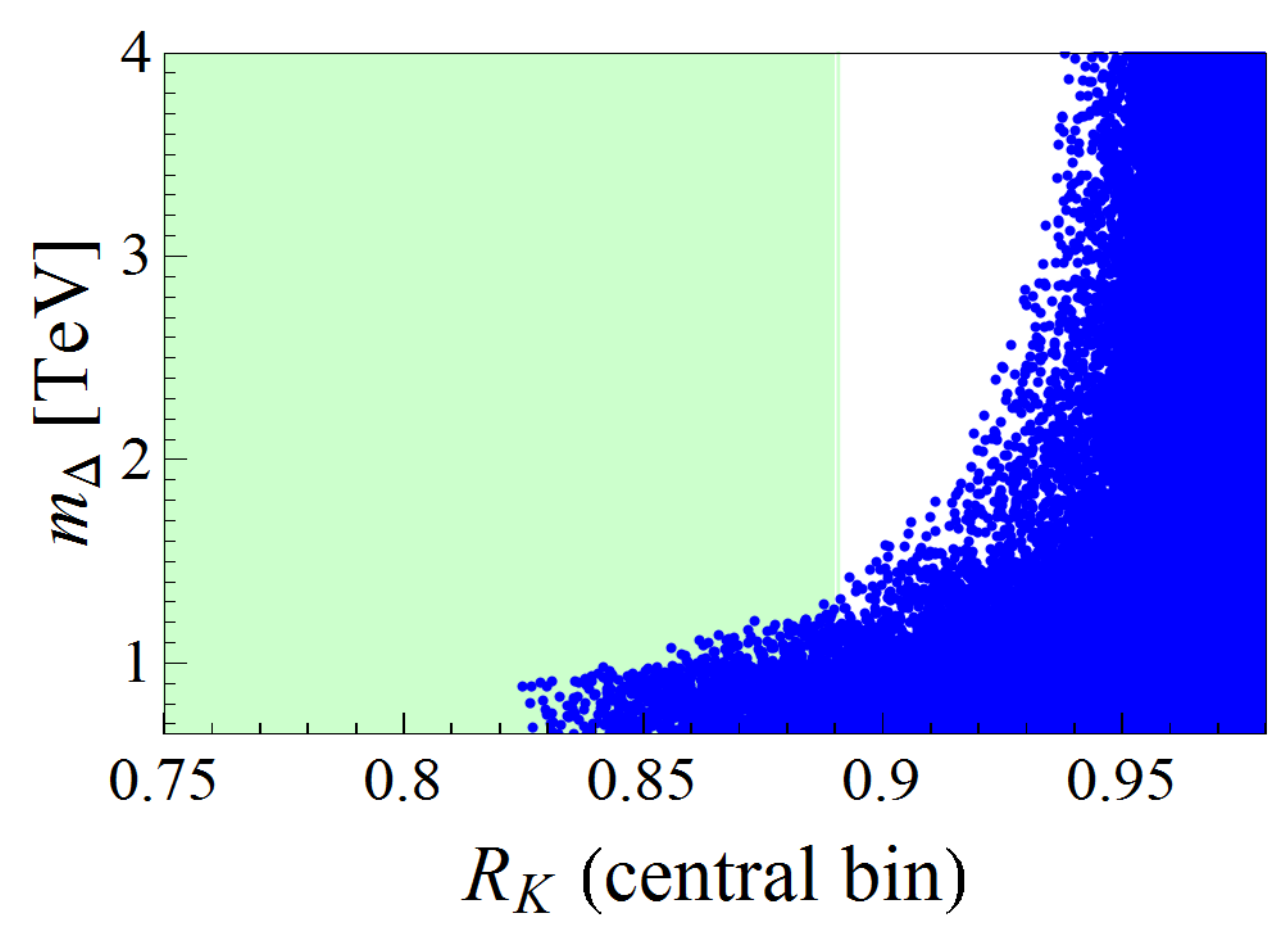}~\includegraphics[width=0.46\linewidth]{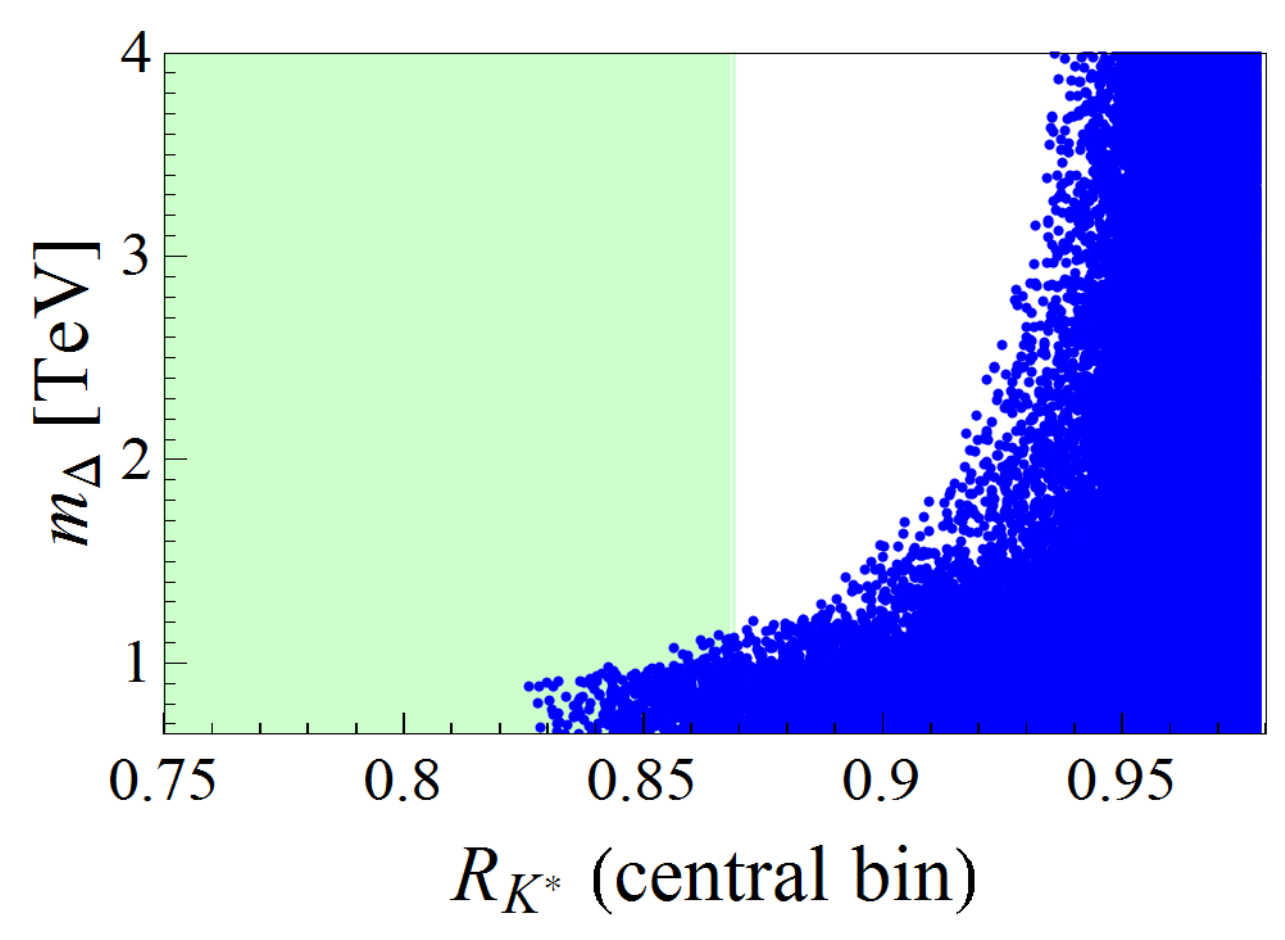}\\
\caption{\small \sl Results of our scan of parameters consistent with all constraints discussed in the previous section in which the leptoquark mass $m_\Delta$ is varied too. We see that the $1.5 \sigma$  consistency requirement with the values of LHCb for $R_K$ and $\Rkst$ in the central $q^2$-bin (shaded area) results in $m_\Delta < 1.2$~TeV. 
}
\label{fig:5}
\end{figure}

As a curiosity we can now proceed the other way around and perform a scan of parameters by leaving $m_\Delta$ as a free parameter, and then check how large one can take $m_\Delta$ and still remain e.g.~$1.5 \sigma$-compatible with $R_K$ and $\Rkst$ reported by LHCb in the central $q^2$-bin. The result of this exercise is shown in Fig.~\ref{fig:5}, from which we see that $m_\Delta < 1.2$~TeV. 
%%%%%%%%%%%%%%%%%%%%%%%%%%%%%%%%%%%%%%%%
%%%%%%%%%%%%%%%%%%%%%%%%%%%%%%%%%%%%%%%%

\

We now enumerate the predictions of this model:
\begin{enumerate}
\item Like we mentioned before, this model does not induce the tree-level or the one-loop contribution to the $B_s-\bar B_s$ mixing amplitude.  
\item Using Eq.~\eqref{eq:C9new} and by taking into account the constraints on the couplings $g_{L,R}^{q\tau}$, we were able to compute $C_9^{\tau\tau}=-C_{10}^{\tau\tau}$ from which 
we computed the branching fractions $\cb (B_s\to \tau\tau)$ and $\cb (B\to K^{(\ast )}\tau\tau)_{{\rm large}\, q^2}$. We obtain, $ -0.46 \leq C_9^{\tau\tau} \leq 0.06$ 
for $m_\Delta = 650$~GeV, which then gives:
\bea
&& 0.78 \leq {\cb (B_s\to \tau\tau)\over \cb (B_s\to \tau\tau)^\mathrm{SM} } \leq 1.03  \,,\nn\\[1.7ex]
&& 0.79\leq {\cb (B\to K \tau\tau)_{q^2\in [15,19]\, \gev^2} \over \cb (B \to K \tau\tau)^\mathrm{SM}_{q^2\in [15,19]\, \gev^2} } \leq 1.03 \,,\nn\\[1.7ex]
&&0.77 \leq {\cb (B\to K^\ast  \tau\tau)_{q^2\in [15,19]\, \gev^2}\over \cb (B \to K^\ast \tau\tau)^\mathrm{SM}_{q^2\in [15,19]\, \gev^2} }  \leq 1.03 \,.
\eea
For $m_\Delta = 1$~TeV, we obtain, $-0.17 \leq C_9^{\tau\tau} \leq 0.03$, which leads to:
\bea
&& 0.92 \leq {\cb (B_s\to \tau\tau)\over \cb (B_s\to \tau\tau)^\mathrm{SM} } \leq 1.01  \,,\nn\\[1.7ex]
&& 0.92\leq {\cb (B\to K \tau\tau)_{q^2\in [15,19]\, \gev^2} \over \cb (B \to K \tau\tau)^\mathrm{SM}_{q^2\in [15,19]\, \gev^2} } \leq 1.01 \,,\nn\\[1.7ex]
&&0.91 \leq {\cb (B\to K^\ast  \tau\tau)_{q^2\in [15,19]\, \gev^2}\over \cb (B \to K^\ast \tau\tau)^\mathrm{SM}_{q^2\in [15,19]\, \gev^2} }  \leq 1.01 \,.
\eea
\item Our model allows for lepton flavor violation, as in most scenarios aiming to explain the LFUV effects~\cite{Glashow:2014iga}.~\footnote{Note, however, that in general a LFUV does not necessarily imply the LFV, as discussed and emphasized in Ref.~\cite{Alonso:2015sja}. 
} 
Again, after inserting the values (intervals) of the couplings $g_L^{\, q\mu}$ and $g_L^{\, q\tau}$ into  Eq.~\eqref{eq:C9new}, we obtain
\bea
\cb (B\to K\mu \tau) \lesssim \left\{ (4.6 \times 10^{-9})_{m_\Delta = 650\,\gev}\,, {(1.5 \times 10^{-9})_{m_\Delta = 1\,\tev}  } \right\}\,,
\eea 
whereas the branching fractions for similar decay modes can be obtained from the ratios which are independent on the Wilson coefficients~\cite{Becirevic:2016zri}:
\begin{equation}\label{eq:ratioS}
\dfrac{\mathcal{B}(B\to K^\ast\mu\tau)}{\mathcal{B}(B\to K\mu\tau)}\approx 1.8 \qquad\quad\mathrm{and}\qquad\quad \dfrac{\mathcal{B}(B_s\to \mu\tau)}{\mathcal{B}(B\to K\mu\tau)}\approx 0.9.
\end{equation}
Since the LFV and lepton flavor conserving modes are related by the same model parameters, there is obviously a correlation between various rates. A typical one is shown in Fig.~\ref{fig:8}, where we see that the LFV mode can be significant even for $\cb (B_s\to \tau\tau)$ perfectly consistent with its Standard Model value.
%%%%%%%%%%%%%%%%%%%%%%%%%%%%%%%%%%%%%%%%
\begin{figure}[t!]
\centering
\includegraphics[width=0.55\linewidth]{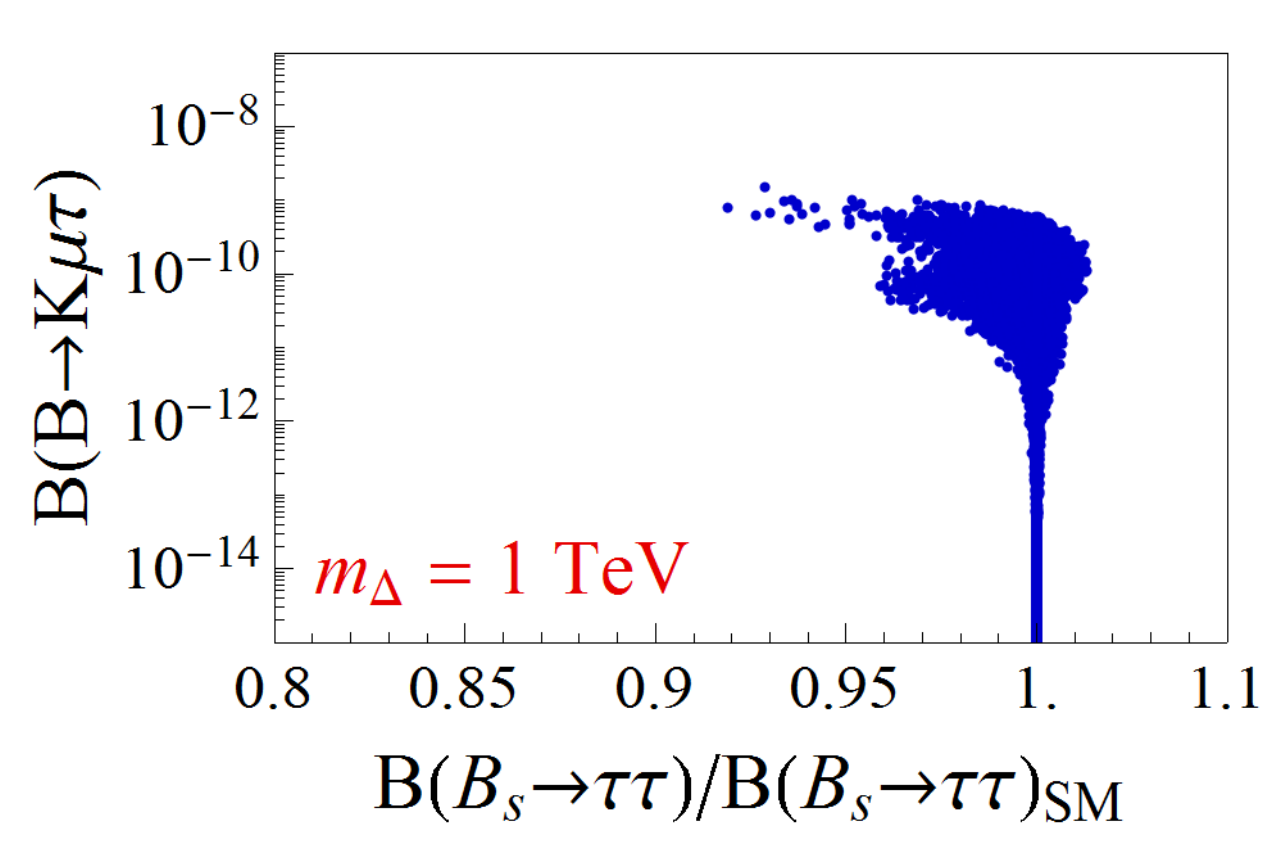}\\
\caption{\small \sl Correlation between $\cb (B_s\to \tau\tau)$ and $\cb (B\to K \mu\tau)$ as obtained in our model. 
}
\label{fig:8}
\end{figure}
%%%%%%%%%%%%%%%%%%%%%%%%%%%%%%%%%%%%%%%%
\item Another interesting LFV mode is $Z\to \mu \tau$. The expression given in Eq.~\eqref{eq:CVL} is trivially extended to the LFV case by simply replacing  
$g_L^{q \ell} g_L^{q\ell \ast }\to g_L^{q \ell_1} g_L^{q\ell_2 \ast }$. We obtain that the maximal allowed values can be quite large, namely,
\bea
\cb (Z\to \mu\tau)  \lesssim \left\{ (4 \times 10^{-7})_{m_\Delta = 650\,\gev}\,, (2.1 \times 10^{-7})_{m_\Delta = 1\,\tev} \right\}\,,
\eea
and could be an opportunity for future experiments.
\item We have checked that our model provides a very small contribution to $\cb (t\to b \tau \nu)$, which is well within the experimental error~\cite{Aaltonen:2014hua}.
\item We also computed the Wilson coefficient relevant to $\cb (B\to K\nu\nu)$ and found that our model can bring only a small reduction with respect to the Standard Model value, i.e. 
\bea
0.94 \leq  {\cb (B\to K\nu\nu)\over \cb (B\to K\nu\nu)^\mathrm{SM}}  \leq  1\,,
\eea
the reduction being more pronounced for smaller leptoquark masses, namely $m_\Delta= 650$~GeV.
\end{enumerate}

\subsection{Consequence on the direct searches of the leptoquark state}

%%%%%%%%%%%%%%%%%%%%%%%%%%%%%%%%%%%%%%%%
%%%%%%%%%%%%%%%%%%%%%%%%%%%%%%%%%%%%%%%%
\begin{figure}[t!]
\centering
\includegraphics[width=0.45\linewidth]{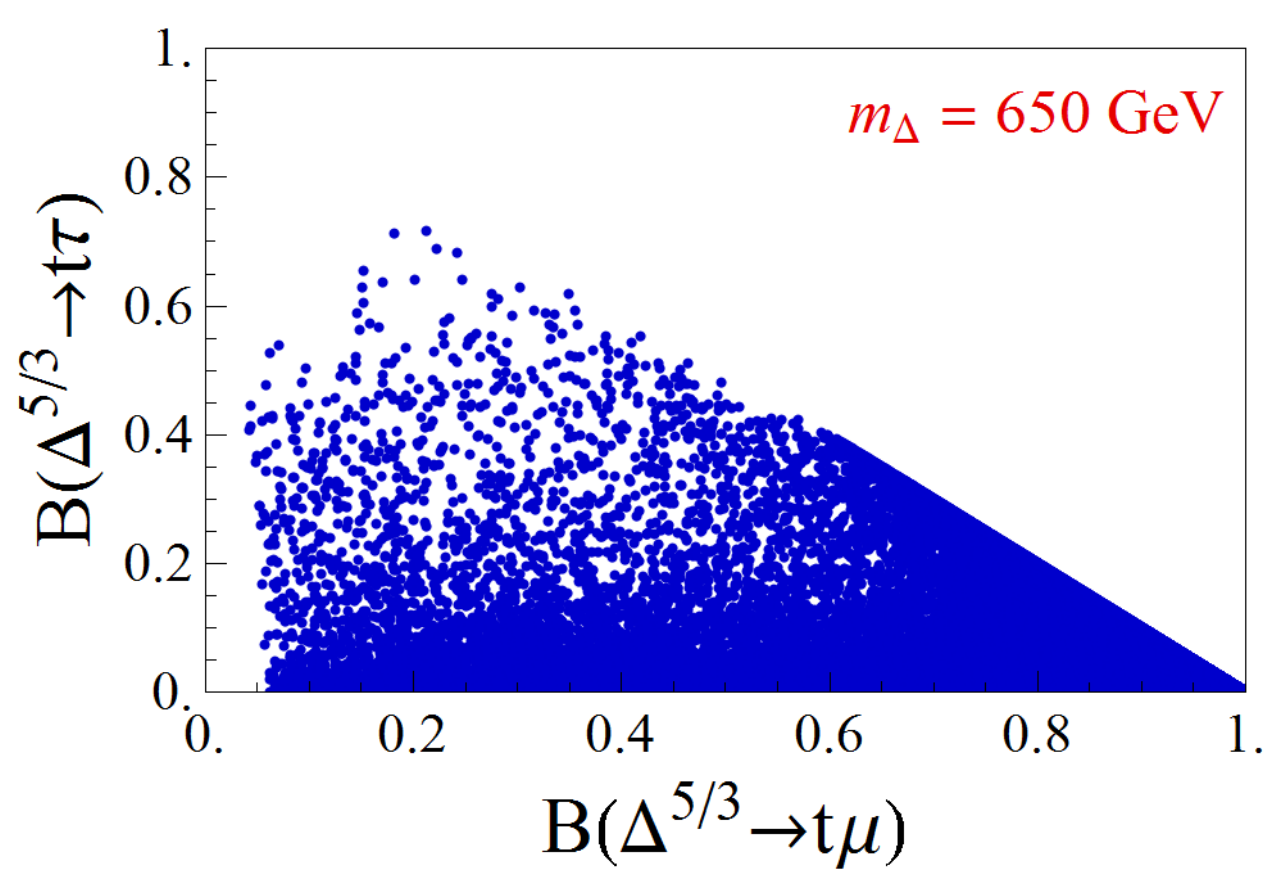}~\includegraphics[width=0.45\linewidth]{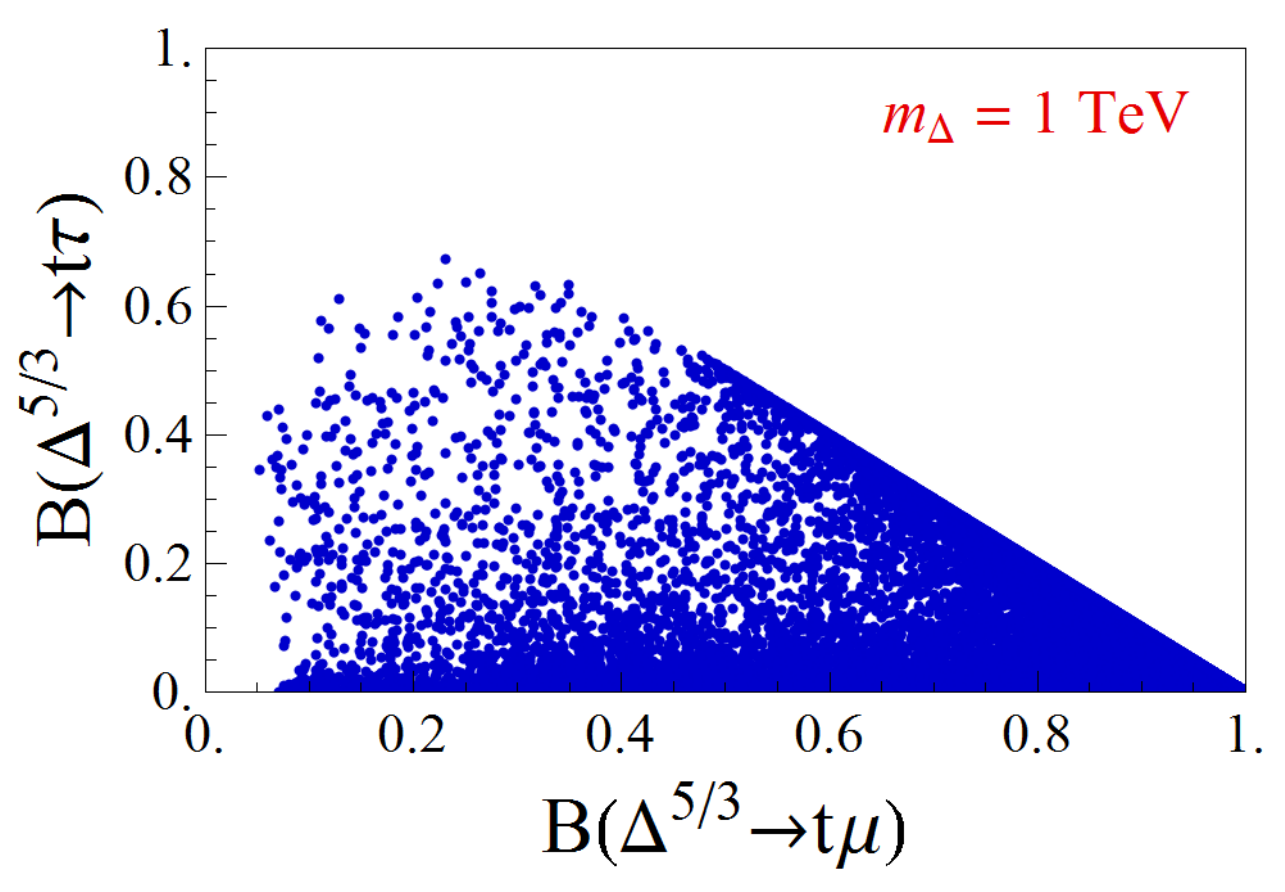}
\caption{\small \sl Branching fractions of the dominant decay modes of $\Delta^{(5/3)}$ as obtained from 
the constraints on the relevant couplings discussed in the body of this paper.}
\label{fig:71}
\end{figure}
%%%%%%%%%%%%%%%%%%%%%%%%%%%%%%%%%%%%%%%%
%%%%%%%%%%%%%%%%%%%%%%%%%%%%%%%%%%%%%%%%
So far we have assumed the value of the leptoquark mass to be either $m_\Delta =650$~GeV or  $m_\Delta =1$~TeV, 
both being consistent with direct searches, cf. Ref.~\cite{direct}. We find that the experimental bound, $m_\Delta \gtrsim 650$~GeV, 
is very conservative and the reason for this can be understood from the assumptions made in the LHC searches. So far the attempts for direct detection of  
the leptoquark states, present in our model, only included the decays
\bea
\Delta^{(2/3)}\to t \nu , \quad\mathrm{and}\quad \Delta^{(5/3)}\to t \tau ,
\eea
for which they assumed $\cb (\Delta^{(2/3)}\to t \nu)=1$, and $\cb (\Delta^{(5/3)}\to t \tau)=1$. 
The resulting bound, $m_\Delta \gtrsim 650$~GeV, would be considerably lower if one also considered 
\bea
\Delta^{(2/3)}\to c \nu , \quad\mathrm{and}\quad \Delta^{(5/3)}\to \ t\mu , \ c\tau, c\mu ,
\eea
and then used the fact that the branching fractions of the above-mentioned modes are less then one. 
With our couplings we can compute the relevant decay rates. We derived the necessary expression for the decay of $\Delta^{(2/3,5/3)}$, namely,
\bea
\Gamma (\Delta^{(2/3)}\to u\, \nu_i) = \Gamma (\Delta^{(5/3)}\to u\, \ell)  = \vert g_L^{\, u i}\vert^2 \frac{ (m_\Delta^2 - m_u^2)^2}{8\pi m_\Delta^3}, 
\eea
where $u\in \{c, t\}$ and $i\in \{\mu, \tau\}$. Notice that we neglect the contribution proportional to $g_R^{b\tau}$ due to its smallness. Furthermore,  
the decay rate $\Gamma (\Delta^{(2/3)}\to b \tau)$ is indeed completely negligible. 
From the above formulas it is then easy to reconstruct the relevant branching fractions 
for the modes searched experimentally. The net result is that the bound on $m_\Delta$ becomes lower. In other words the values we use, $m_\Delta \geq 650$~GeV, are in fact very conservative.   
Note also that the modes with the charm quark are experimentally very challenging at the LHC.

In Fig.~\ref{fig:71} we show the possible values for $\cb ( \Delta^{(5/3)} \to t\tau )$ and $\cb ( \Delta^{(5/3)}  \to  t\mu )$, consistent with all the constraints discussed in Sec.~\ref{sec:constr}. 
This information can be used in the forthcoming attempts at LHC to detect the leptoquark through $\Delta^{(5/3)}  \to t \mu$ channel.

\section{Summary}
\label{sec:conc}
%%%%%%%%%%%%%%%%%%%%%%%%%%%%%%%%%%
%%%%%%%%%%%%%%%%%%%%%%%%%%%%%%%%%%
In this paper we discussed a peculiar form of the $R_2$ model, i.e.~a model in which one postulates the existence of the low energy doublet of mass degenerate scalar leptoquarks 
with hypercharge $Y=7/6$. 
A peculiarity of the model lies in the fact that the couplings of leptoquarks to $s$-quark are forbidden which then means that a contribution of the model to the $b\to s\mu\mu$ decay modes 
is induced by a loop, cf. Fig~\ref{fig:1}. We computed the relevant Wilson coefficients which, in this model, satisfy a condition $C_9^{\mu\mu}=-C_{10}^{\mu\mu}$ and can explain 
the experimental hints on the LFUV in the exclusive $B\to K^{(\ast )}\mu\mu$ modes recently reported by the LHCb collaboration [cf. Eqs.~\eqref{exp:RK} and \eqref{exp:RKstar}]. 

Since the model satisfies $C_9^{\mu\mu}=-C_{10}^{\mu\mu}$, we first showed that from the constraints on the only new physics contribution ($C_9^{\mu\mu}$), deduced from comparison of the expressions derived in effective field theory with quantities for which the hadronic uncertainties are under good theoretical control, one can indeed show that both $R_K<R_K^\mathrm{SM}$ and $\Rkst < R_{K^\ast}^\mathrm{SM}$. 

We then impose a number of constraints on our model and show that the agreement with experiment remains very good. We find that our values of $R_K$ and $R_{K^\ast}$ agree with experiment to $1.1\sigma$ in the central $q^2$-bin, while the agreement with $R_{K^\ast}^\mathrm{exp}$ in the low $q^2$-bin is only at the level of $1.8\sigma$. 
In this discussion, and in agreement with direct searches, we examined the situations by fixing the leptoquark mass to either $m_\Delta =650\ \gev$ or to $m_\Delta =1\ \tev$, and in both cases the agreement remains as indicated above.

We then discussed several predictions of this model, which include a rather large upper bound on the LFV mode, $\cb(Z\to\mu\tau) \lesssim \mathcal{O}(10^{-7})$, while the 
one on $\mathcal{B}(B\to K\mu\tau)$ and similar decay modes is $\mathcal{O}(10^{-9})$.

We also argued that the assumptions used to derive the lower bound on the leptoquark mass from the direct searches can be reinterpreted if we constrain the Yukawa couplings and compute the decay rates to dominant decay channels. In that way the branching fractions can be bounded and the values of the lower bound to the leptoquark mass lowered.

\vskip 1.4cm
\noindent 
{\bf \large Acknowledgments:} {\sl We would like to thank Rafael Lopes Coelho de S\'a and Svjetlana Fajfer for discussions. This project has received funding from the European Union's Horizon 2020 research and innovation program under the Marie Sklodowska-Curie grant agreements No. 690575 and No. 674896.  }

\end{document}